# Protein Dynamics Beyond Structure Prediction

**Juliette Griffié[1], Sviatlana Shashkova[2], Antonio Ciarlo[2], Sreekanth K. Manikandan[2], Claes Andréasson[3], Malin Bäckström[4], Tristan Bereau[5], Hjalmar Brismar[6,7], Carlos Bustamante[8-15], Marta Carroni[16], Roberto Covino[17], Andreas Dahlin[18], Sebastian Deindl[19,20], Lucie Delemotte[6], Arne Elofsson[16], John Eriksson[21], Giovanna Fragneto[22], Anders Gunnarsson[23], Per Hammarström[24,25], Caroline Ingre[26], Christian Kaiser[27], Petronella Kettunen[28,29], Mark C. Leake[30], Benjamin Loos[31], Anna Månberg[32], Antonia S. J. S. Mey[33], Richard Neutze[34,†], Thomas Nyström[35], Karl Palmås[36], Charley Schaefer[37], Markus J. Tamás[34], Nicola Ticozzi[38,39], Tomás S. Pilvelic[40], Jacopo Sacquegno[41], B.M. (Betty) Tijms[42,43], Gunnar von Heijne[1], Björn Wallner[24,25], Vitali Zhaunerchyk[2], Simon Olsson[44], Joana B. Pereira[26], Julia Fernandez-Rodriguez[45,46], Fredrik Westerlund[47], Giovanni Volpe[2,48]**

1. Department of Biochemistry and Biophysics, Stockholm University, Stockholm, Sweden
2. Department of Physics, University of Gothenburg, 41296 Gothenburg, Sweden
3. Department of Molecular Biosciences, The Wenner-Gren Institute, Stockholm University, S-106 91 Stockholm, Sweden
4. Mammalian Protein Expression Core facility and Protein Production Sweden, Sahlgrenska Academy, Medicinaregatan 9E, University of Gothenburg, Gothenburg, Sweden
5. Institute for Theoretical Physics, Heidelberg University, 69120 Heidelberg, Germany. 2. Interdisciplinary Center for Scientific Computing (IWR), Heidelberg University, 69120 Heidelberg, Germany
6. Science for Life Laboratory, Department of Applied Physics, KTH, Sweden
7. Science for Life Laboratory, Department of Women's and Children's Health, Karolinska Intitutet, Solna, Sweden
8. Biophysics Graduate Group, University of California, Berkeley, CA, USA
9. Jason L. Choy Laboratory of Single Molecule Biophysics, University of California, Berkeley, CA, USA
10. California Institute for Quantitative Biosciences-QB3, University of California, Berkeley, CA, USA
11. Howard Hughes Medical Institute, University of California, Berkeley, CA, USA
12. Department of Chemistry, University of California, Berkeley, CA, USA
13. Department of Physics, University of California, Berkeley, CA, USA
14. Department of Molecular & Cell Biology, University of California, Berkeley, CA, USA
15. Kavli Energy Nanosciences Institute, University of California, Berkeley, CA, USA
16. Science for Life Laboratory, Department of Biochemistry and Biophysics, Stockholm University, Stockholm, Sweden
17. Institute of Computer Science, Goethe University Frankfurt & Frankfurt Institute for Advanced Studies, Frankfurt am Main, Germany
18. Department of Chemistry and Chemical Engineering, Chalmers University of Technology, Gothenburg, Sweden
19. Department of Cell and Molecular Biology, Science for Life Laboratory, Uppsala University, Uppsala, 75105, Sweden
20. Interfaculty Institute of Biochemistry (IFIB), University of Tübingen, Auf der Morgenstelle 34, Tübingen, 72076, Germany
21. Eurobioimaging Hub Office, Turku, Finland
22. European Spallation Source ERIC, P.O. Box 176, SE-221 00 Lund, Sweden and Lund University, Physical Chemistry P.O. Box 124, SE-221 00 Lund

23. Protein, Structure and Biophysics, Discovery Sciences, BioPharmaceuticals R&D, AstraZeneca, Gothenburg 431 83, Sweden
24. Department of Physics, Chemistry and Biology (IFM), Linköping University, Linköping, 581 83, Sweden
25. Science for Life Laboratory, Linköping University, Linköping, Sweden
26. Neuro Division, Department of Clinical Neuroscience, Karolinska Institute, Sweden
27. Bijvoet Centre for Biomolecular Research, Department of Chemistry, Utrecht University, The Netherlands
28. Department of Psychiatry and Neurochemistry, Institute of Neuroscience and Physiology, Sahlgrenska Academy, University of Gothenburg, Gothenburg, Sweden
29. Department of Neuropsychiatry, Sahlgrenska University Hospital, Region Västra Götaland, Mölndal, Sweden
30. School of Physics, Engineering and Technology and Department of Biology, University of York, York, YO10 5DD, United Kingdom
31. Department of Physiological Sciences, Stellenbosch University, Stellenbosch, 7600, South Africa
32. Department of Protein Science, KTH Royal Institute of Technology, SciLifeLab, Stockholm, Sweden
33. EaStCHEM School of Chemistry, University of Edinburgh, David Brewster Road, Edinburgh, EH9 3FJ, United Kingdom
34. Department of Chemistry and Molecular Biology, University of Gothenburg, S-405 30 Göteborg, Sweden
35. Department of Medical Biochemistry and Cell Biology, Institute of Biomedicine, Sahlgrenska Academy, University of Gothenburg, Sweden
36. Division of Science, Technology and Society, Chalmers University of Technology, 412 96 Gothenburg, Sweden
37. School of Physics and Astronomy & Astbury Centre for Structural Molecular Biology, University of Leeds, Woodhouse Lane, Leeds, LS2 9JT, United Kingdom
38. Department of Neurology, IRCCS Istituto Auxologico Italiano, 20149 Milan, Italy
39. Department of Pathophysiology and Transplantation, Università degli Studi di Milano,20122 Milan, Italy
40. MAXIV Laboratory, Lund University. Fotongatan 2 224 84 Lund Sweden
41. Independent Visual Science Communicator, Milan, Italy.
42. Alzheimer Center Amsterdam, Department of Neurology, Amsterdam UMC, 1081 HZ, Amsterdam, The Netherlands
43. Amsterdam Neuroscience, Neurodegeneration, Amsterdam, The Netherlands
44. Department of Computer Science and Engineering, Chalmers University of Technology and University of Gothenburg, SE-41296 Gothenburg, Sweden
45. Centre for Cellular imaging, Sahlgrenska Academy, University of Gothenburg, Medicinaregatan 5A, 405 30 Gothenburg, Sweden
46. Integrated Microscopy Technologies, Science for Life Laboratory, University of Gothenburg, Sweden
47. Department of Life Sciences, Chalmers University of Technology, 412 96 Gothenburg, Sweden
48. Science for Life Laboratory, Department of Physics, University of Gothenburg, 41296 Gothenburg, Sweden

† Deceased

**The ability to predict protein three-dimensional structures from amino acid sequences represents a landmark achievement in molecular biology, where recent deep learning approaches such as AlphaFold are the culmination of decades of work. Yet, the quantitative understanding of how protein sequences give rise to dynamic conformational changes and higher-order assemblies remains unsolved. Structure prediction often provides an accurate model of a stable or functional conformational state, but it does not specify the pathway, kinetics, or mechanisms that drive polypeptide chains to their native states, interconvert between functional conformations, assemble into macromolecular complexes, or misfold into pathological conformations. Folding and conformational states are dynamic, stochastic processes, shaped by sequence, energy, co-translational constraints, chaperone machineries, and the physicochemical conditions of the cellular environment. Recent advances now position the field to move beyond static structural endpoints toward a mechanistic understanding of folding dynamics in living systems. Single-molecule techniques enable time-resolved observation of folding trajectories and intermediate states hitherto hidden by traditional structural biology approaches, while computational innovations and data-driven approaches offer new ways to integrate heterogeneous data across scales. Together, these developments create an opportunity to establish a predictive framework linking sequence to folding pathways, conformational landscapes, assembly processes, misfolding propensities, and cellular outcomes. In this Roadmap, we review the current conceptual landscape of protein folding, examine the experimental and theoretical gaps that remain, and discuss emerging strategies that integrate high-resolution measurements with multiscale modeling. We outline a roadmap toward a quantitative and predictive science of protein folding dynamics, conformational kinetics, and macromolecular self-assembly. Realizing this vision would transform our understanding of the dynamics of molecular self-organization, from the folding of individual polypeptides to the emergence of dynamic macromolecular complexes. In turn, this will enable rational control of folding and misfolding in health and disease, extend protein engineering principles beyond static structural design, and establish a mechanistic foundation for predictive and personalized interventions in proteostasis-related disorders.**

## Introduction

More than six decades after Christian Anfinsen proposed that the native structure of a protein is encoded in its amino acid sequence,[1] the problem of how protein sequences encode conformational dynamics and self-assembly into functional macromolecular complexes remains only partially solved. Anfinsen's thermodynamic hypothesis established a foundational principle: under appropriate conditions, a polypeptide chain contains within its sequence the information required to reach its native three-dimensional (3D) structure. That insight transformed molecular biology. Yet, it eventually raised deeper questions that remain unresolved today: How does a protein navigate the vast space of possible conformations to reach its folded state in biologically relevant timescales (*Levinthal's paradox*[2])? How are folding pathways encoded in the amino acid sequence? What determines folding kinetics, intermediate states, and misfolding routes?

Over the decades, experimental and theoretical work has provided important pieces of the puzzle. Conceptually, two major frameworks have guided thinking about folding mechanisms. *Energy landscape theory* describes folding as diffusion over a multidimensional, rugged free-energy surface shaped like a funnel toward the native state.[3] In contrast, hierarchical or *foldon models* emphasize the sequential stabilization of cooperative structural units along partially ordered pathways.[4] These perspectives are not mutually exclusive, but neither has yielded a general, predictive mapping from sequence to folding trajectory. Protein conformations span a continuum from ordered and functional to disordered and denatured states, with some proteins switching folds without sequence changes[5] and others functioning as dynamic ensembles[6] that can drive processes such as liquid–liquid phase separation.[7] Some proteins can adopt alternative aggregated structures, such as *amyloids*, which may be functional and/or linked to disease, and in some cases these structures can self-propagate between molecules.[8] In this broader view, protein folding can be understood as a paradigmatic instance of a more general problem: how sequence encodes dynamic conformational landscapes that enable function, interaction, and self-assembly.

Experimentally, an expansive range of techniques has been developed to probe protein structure and dynamics (Table 1). Ensemble approaches, such as circular dichroism (CD),[9]

---

[1] C. B. Anfinsen, "Principles That Govern the Folding of Protein Chains," *Science* 181 (1973): 223–30, https://doi.org/10.1126/science.181.4096.223.
[2] C. Levinthal, "Are There Pathways for Protein Folding?," *Journal de Chimie Physique et de Physico-Chimie Biologique* 65 (1968): 44–45, https://doi.org/10.1051/jcp/1968650044.
[3] J. D. Bryngelson et al., "Funnels, Pathways, and the Energy Landscape of Protein Folding: A Synthesis," *Proteins: Structure, Function, and Bioinformatics* 21 (1995): 167–95, https://doi.org/10.1002/prot.340210302.
[4] S. W. Englander and L. Mayne, "The Nature of Protein Folding Pathways," *Proceedings of the National Academy of Sciences* 111 (2014): 15873–80, https://doi.org/10.1073/pnas.1411798111.
[5] A. G. Murzin, "Metamorphic proteins," *Science* 320 (2008): 1725–26, https://doi.org/10.1126/science.1158868.
[6] V. J. Hilser et al., "Statistical Thermodynamics of the Protein Ensemble: Mediating Function and Evolution," *Annual Review of Biophysics* 54 (2025): 227–47, https://doi.org/10.1146/annurev-biophys-061824-104900.
[7] Y. Shin and C. P. Brangwynne, "Liquid Phase Condensation in Cell Physiology and Disease," *Science* 357 (2017): eaaf4382, https://doi.org/10.1126/science.aaf4382.
[8] P. Kulkarni et al., "Evolving Concepts of the Protein Universe," *iScience* 28 (2025): 112012, https://doi.org/10.1016/j.isci.2025.112012.
[9] N. J. Greenfield, "Using Circular Dichroism Spectra to Estimate Protein Secondary Structure," *Nature Protocols* 1 (2006): 2876–90, https://doi.org/10.1038/nprot.2006.202.

Fourier transform infrared spectroscopy (FTIR),[10] stopped-flow kinetics,[11] nuclear magnetic resonance (NMR) spectroscopy,[12] and hydrogen–deuterium exchange mass spectrometry (HDX-MS)[13,14] have revealed key insights into stability and intermediate states. Structural approaches, including X-ray crystallography,[15] X-ray solution scattering (XRS, SAXS, WAXS),[16, 17] and cryo-electron microscopy (cryo-EM),[18] have defined atomic architectures with extraordinary precision. More recently, single-molecule methods, particularly based on single-molecule fluorescence[19] and force spectroscopy,[20] have begun to expose folding trajectories and stochastic heterogeneity directly. At the cellular and organismal levels, advanced imaging,[21] genetic perturbations,[22] and disease models[23] have enabled mechanistic investigation of protein misfolding and aggregation within their native physiological context. Yet, each of these approaches accesses only part of the spatial and temporal spectrum of folding and protein dynamics.

Computational approaches have evolved in parallel (Table 2). Early lattice models[24] and simplified energy functions[25] provided conceptual clarity by demonstrating how minimal interaction rules can drive chain compaction and generate sequence-dependent folded structures. All-atom molecular dynamics simulations now offer atomistic views of folding events for selected systems,[26] aided by enhanced sampling techniques[27] and specialized hardware.[28] Coarse-grained models extend accessible length and timescales, enabling exploration of larger assemblies.[29] For folded proteins, AlphaFold2[30] and RoseTTAFold[31] transformed structure prediction. In parallel, coarse-grained and data-informed models such as CALVADOS[32,33] have improved modelling of intrinsically disordered proteins and phase behaviour. Crucially, these developments did not emerge in a vacuum: they are the result of decades of structural biology research and the accumulation of large, curated datasets in the Protein Data Bank,[34] built largely from X-ray crystallography and later complemented by NMR spectroscopy and cryo-EM, and protein sequence data made possible by next-generation sequencing technology and data-sharing platforms such as UniProt.[35] The success of AlphaFold2 and RoseTTAFold is therefore as much a testament to experimental infrastructure and data sharing as it is to algorithmic innovation.

---

[10] A. Barth, "Infrared spectroscopy of proteins," *Biochimica et Biophysica Acta (BBA)-Bioenergetics* 1767 (2007): 1073–101, https://doi.org/10.1016/j.bbabio.2007.06.004.
[11] A. Esadze and J. Iwahara, "Stopped-Flow Fluorescence Kinetic Study of Protein Sliding and Intersegment Transfer in the Target DNA Search Process," *Journal of Molecular Biology* 426 (2014): 230–44, https://doi.org/10.1016/j.jmb.2013.09.019.
[12] A. Bax and S. Grzesiek, "Methodological Advances in Protein NMR," *Accounts of Chemical Research* 26 (1993): 131–38, https://doi.org/10.1021/ar00028a001.
[13] S. W. Englander, "Hydrogen Exchange and Mass Spectrometry: A Historical Perspective," *Journal of the American Society for Mass Spectrometry* 17 (2006): 1481–89, https://doi.org/10.1016/j.jasms.2006.06.006.
[14] A. J. Percy et al., "Probing Protein Interactions with Hydrogen/Deuterium Exchange and Mass Spectrometry—a Review," *Analytica Chimica Acta* 721 (2012): 7–21, https://doi.org/10.1016/j.aca.2012.01.037.
[15] J. C. Kendrew et al., "A Three-Dimensional Model of the Myoglobin Molecule Obtained by x-Ray Analysis," *Nature* 181 (1958): 662–66, https://doi.org/10.1038/181662a0.
[16] H. S. Cho et al., "Time-Resolved X-Ray Scattering Studies of Proteins," *Current Opinion in Structural Biology* 70 (2021): 99–107, https://doi.org/10.1016/j.sbi.2021.05.002.
[17] S. V. Kathuria et al., "Microsecond Barrier-Limited Chain Collapse Observed by Time-Resolved FRET and SAXS," *Journal of Molecular Biology* 426 (2014): 1980–94, https://doi.org/10.1016/j.jmb.2014.02.020.
[18] A. Amunts et al., "Structure of the Yeast Mitochondrial Large Ribosomal Subunit," *Science* 343 (2014): 1485–89, https://doi.org/10.1126/science.1249410.
[19] D. Nettels et al., "Single-molecule FRET for probing nanoscale biomolecular dynamics," *Nature Reviews Physics* 6 (2024): 587–605, https://doi.org/10.1038/s42254-024-00748-7.
[20] C. Bustamante et al., "Mechanical Processes in Biochemistry," *Annual Review of Biochemistry* 73 (2004): 705–48, https://doi.org/10.1146/annurev.biochem.72.121801.161542.
[21] E. Betzig et al., "Imaging Intracellular Fluorescent Proteins at Nanometer Resolution," *Science* 313 (2006): 1642–45, https://doi.org/10.1126/science.1127344.
[22] O. Shalem et al., "Genome-Scale CRISPR-Cas9 Knockout Screening in Human Cells," *Science* 343 (2014): 84–87, https://doi.org/10.1126/science.1247005.
[23] C. Soto and S. Pritzkow, "Protein Misfolding, Aggregation, and Conformational Strains in Neurodegenerative Diseases," *Nature Neuroscience* 21 (2018): 1332–40, https://doi.org/10.1038/s41593-018-0235-9.
[24] K. Yue et al., "A Test of Lattice Protein Folding Algorithms," *Proceedings of the National Academy of Sciences* 92 (1995): 325–29, https://doi.org/10.1073/pnas.92.1.325.
[25] K. A. Dill, "Theory for the Folding and Stability of Globular Proteins," *Biochemistry* 24 (1985): 1501–9, https://doi.org/10.1021/bi00327a032.

However, accurate prediction of static structure does not equate to solving the protein folding problem.[36] Structure prediction provides the endpoints of folding under idealized conditions; it does not provide the pathway, the kinetics, or the dynamic ensemble through which that endpoint is reached. It does not describe how folding competes with aggregation, how molecular chaperones remodel intermediates, how proteins interconvert between functional conformations, assemble into macromolecular complexes, or how membrane proteins fold and insert themselves within lipid membranes, or how mutations alter kinetic partitioning between native and misfolded states. In this sense, the remarkable success of structure prediction risks narrowing our focus to the final state, while the mechanistic core of folding, its dynamics in time, remains incompletely understood.

---

[26] S. J. Weiner et al., "An All Atom Force Field for Simulations of Proteins and Nucleic Acids," *Journal of Computational Chemistry* 7 (1986): 230–52, https://doi.org/10.1002/jcc.540070216.
[27] J. Hénin et al., "Enhanced Sampling Methods for Molecular Dynamics Simulations [Article v1.0]," *Living Journal of Computational Molecular Science* 4 (2022): 1583, https://doi.org/10.33011/livecoms.4.1.1583.
[28] D. E. Shaw et al., "Anton 2: Raising the Bar for Performance and Programmability in a Special-Purpose Molecular Dynamics Supercomputer," *SC'14: Proceedings of the International Conference for High Performance Computing, Networking, Storage and Analysis*, November 2014, 41–53, https://doi.org/10.1109/SC.2014.9.
[29] C. Clementi, "Coarse-Grained Models of Protein Folding: Toy Models or Predictive Tools?," *Current Opinion in Structural Biology* 18 (2008): 10–15, https://doi.org/10.1016/j.sbi.2007.10.005.
[30] J. Jumper et al., "Highly Accurate Protein Structure Prediction with AlphaFold," *Nature* 596 (2021): 583–89, https://doi.org/10.1038/s41586-021-03819-2.
[31] M. Baek et al., "Accurate Prediction of Protein Structures and Interactions Using a Three-Track Neural Network," *Science* 373 (2021): 871–76, https://doi.org/10.1126/science.abj8754.
[32] G. Tesei et al., "Accurate Model of Liquid-Liquid Phase Behavior of Intrinsically Disordered Proteins from Optimization of Single-Chain Properties," *Proceedings of the National Academy of Sciences* 118 (2021): e2111696118, https://doi.org/10.1073/pnas.2111696118.
[33] Giulio Tesei and Kresten Lindorff-Larsen, "Improved Predictions of Phase Behaviour of Intrinsically Disordered Proteins by Tuning the Interaction Range," *Open Research Europe* 2 (January 2023): 94, https://doi.org/10.12688/openreseurope.14967.2.
[34] S. K. Burley et al., "RCSB Protein Data Bank: Powerful New Tools for Exploring 3D Structures of Biological Macromolecules for Basic and Applied Research and Education in Fundamental Biology, Biomedicine, Biotechnology, Bioengineering and Energy Sciences," *Nucleic Acids Research* 49 (2021): 437–51, https://doi.org/10.1093/nar/gkaa1038.
[35] The UniProt Consortium, "UniProt: The Universal Protein Knowledgebase in 2023," *Nucleic Acids Research* 51 (2023): 523–31, https://doi.org/10.1093/nar/gkac1052.
[36] K. A. Dill and J. L. MacCallum, "The Protein-Folding Problem, 50 Years On," *Science* 338 (2012): 1042–46, https://doi.org/10.1126/science.1219021.

A central reason for this gap is the separation of scales.[37] In space, folding depends on Ångström-level interactions that stabilize secondary and tertiary structures, nanometer-scale rearrangements that organize domains, and larger-scale interactions with chaperones, membranes, and macromolecular assemblies within the crowded cellular environment on micrometer scales. In time, local backbone fluctuations and side-chain movements occur on nanosecond to microsecond scales, barrier-crossing transitions that define folding pathways typically unfold over milliseconds to seconds, and in pathological contexts, misfolding and aggregation can accumulate over years to decades. It is difficult to capture these scales simultaneously (but when they do, for example, for fast-folding proteins, it has been shown that computational methods are in good agreement with experimental data[38]). Experimental techniques are typically optimized either for high spatial resolution or for physiological context, rarely for both simultaneously; single-molecule techniques resolve kinetic intermediates, yet often operate outside the full cellular environment, whereas cellular and organismal models capture biological complexity, but, even when combined with single-molecule techniques, obscure molecular mechanisms. However, this is the environment where folding occurs and interactions between unfolded proteins and other molecules are prevented as part of the folding process. Moreover, single-molecule methods ignore the vectorial character of co-translational protein folding processes in which proteins fold when they emerge. Similarly, computational methods face trade-offs between atomistic detail and accessible timescales. Despite major advances, there is still no unified framework that seamlessly connects single-molecule folding events to cellular proteostasis and organismal phenotypes.

The field stands at a critical inflection point. On the experimental front, advances in single-molecule force spectroscopy including optical tweezers,[20] [39] [40] high-resolution fluorescence methods[19], and ensemble techniques such as HDX-MS,[41] now enable increasingly precise, time-resolved characterization of folding trajectories at single-molecule resolution. On the computational side, machine learning has demonstrated its power to extract complex mappings from large datasets and to integrate heterogeneous information across scales.[42]

The convergence of high-resolution single-molecule experiments, scalable data acquisition, and AI-driven modeling creates an opportunity to move beyond static structures toward a predictive science of folding pathways, conformational dynamics, and macromolecular self-assembly. Such a framework would not only deepen our understanding of how proteins attain their functional states, but also illuminate why they misfold, how protein quality control networks modulate these processes, and how sequence variations translate into altered folding behavior in health and disease.

---

[37] H. Frauenfelder et al., "The Energy Landscapes and Motions of Proteins," *Science* 254 (1991): 1598–603, https://doi.org/10.1126/science.1749933.
[38] K. Lindorff-Larsen et al., "How Fast-Folding Proteins Fold," *Science* 334 (2011): 517–20, https://doi.org/10.1126/science.1208351.
[39] K. C. Neuman and A. Nagy, "Single-Molecule Force Spectroscopy: Optical Tweezers, Magnetic Tweezers and Atomic Force Microscopy," *Nature Methods* 5 (2008): 491–505, https://doi.org/10.1038/nmeth.1218.
[40] Robert P. Sosa et al., "The Rossmann2×2 Fold Attains Its Native Structure Via a Defined Pathway of Sequential and Cooperative Folding Units," preprint, Biophysics, May 22, 2026, https://doi.org/10.64898/2026.05.21.726993.
[41] L. Konermann et al., "Hydrogen Exchange Mass Spectrometry for Studying Protein Structure and Dynamics," *Chemical Society Reviews* 40 (2011): 1224–34, https://doi.org/10.1039/C0CS00113A.
[42] Frank Noé et al., "Machine Learning for Molecular Simulation," *Annual Review of Physical Chemistry* 71 (2020): 361–90, https://doi.org/10.1146/annurev-physchem-042018-052331.

The clinical urgency of these insights is evident, for example, in neurodegenerative disorders. In Alzheimer's disease, abnormal aggregation and accumulation of amyloid-β (Aβ) and tau amyloid fibrils in the brain can begin 20 years before the onset of cognitive symptoms.[43, 44] Yet, biomarker positivity does not guarantee clinical progression, and many individuals with detectable pathological proteins remain cognitively stable for extended periods.[45] Recent evidence using single molecule and multimodal imaging techniques highlights the complexity of intra-and extracellular protein aggregation[46], and the distinct protein cargo recruitment through the molecular machinery of protein degradation.[47] Specifically, the role of autophagy activity[48], and its engagement with mitochondrial quality control and lysosomal acidification status has received increasingly attention.[49] In amyotrophic lateral sclerosis (ALS), disease progression is highly heterogeneous: onset location, rate of decline, cognitive involvement, and survival vary dramatically among patients.[50] Beyond neurodegeneration, systemic and localized amyloidoses are highly detrimental diseases caused by the buildup of fibrils from misfolded amyloid proteins in most organs of the body depending on the protein and disease.[51] Folding defects also underlie metabolic and secretory disorders, including misfolded insulin or proinsulin variants and cystic fibrosis transmembrane conductance regulator (CFTR) misfolding in cystic fibrosis, where impaired conformational maturation disrupts trafficking and function.[52, 53] Collectively, these observations underscore a central limitation of current approaches: detecting the presence of misfolded or aggregated proteins is not sufficient to predict disease trajectory.[54]

---

[43] Jianping Jia et al., "Biomarker Changes during 20 Years Preceding Alzheimer's Disease," *New England Journal of Medicine* 390 (February 2024): 712–22, https://doi.org/10.1056/NEJMoa2310168.
[44] Yan Li et al., "Timing of Biomarker Changes in Sporadic Alzheimer's Disease in Estimated Years from Symptom Onset," *Annals of Neurology* 95 (May 2024): 951–65, https://doi.org/10.1002/ana.26891.
[45] Anna-Chloé Balageas et al., "Case Report: The Alzheimer's Paradox: A Clinically Stable Amnestic Syndrome with Full Biomarker Positivity and Minimal Imaging Evidence," *Frontiers in Medicine* 12 (August 2025): 1653232, https://doi.org/10.3389/fmed.2025.1653232.
[46] T. Knowles et al., "The Amyloid State and Its Association with Protein Misfolding Diseases," *Nature Reviews Molecular Cell Biology* 15 (2014): 384–96, https://doi.org/10.1038/nrm3810.
[47] D. Lumkwana et al., "Investigating the Role of Spermidine in a Model System of Alzheimer's Disease Using Correlative Microscopy and Super-Resolution Techniques," *Front. Cell Dev. Biol* 10, no. 819571 (2022), https://doi.org/10.3389/fcell.2022.819571.
[48] I. Dikic and Z. Elazar, "Mechanism and medical implications of mammalian autophagy," *Nature Reviews Molecular Cell Biology* 19 (2018): 349–64, https://doi.org/10.1038/s41580-018-0003-4.
[49] A. du Toit et al., "Measuring autophagosome flux," *Autophagy* 14 (2018): 1060–71, https://doi.org/10.1080/15548627.2018.1469590.
[50] P. Masrori and P. Damme, "Amyotrophic Lateral Sclerosis: A Clinical Review," *European Journal of Neurology* 27 (2020): 1918–29, https://doi.org/10.1111/ene.14393.
[51] G. Merlini et al., "Amyloid Nomenclature 2024: Update, Novel Proteins, and Recommendations by the International Society of Amyloidosis (ISA) Nomenclature Committee," *Amyloid* 31 (2024): 249–56, https://doi.org/10.1080/13506129.2024.2405948.
[52] E. F. McDonald et al., "CFTR Folding: From Structure and Proteostasis to Cystic Fibrosis Personalized Medicine," *ACS Chemical Biology* 18 (2023): 2128–43, https://doi.org/10.1021/acschembio.3c00310.
[53] A. Arunagiri et al., "Proinsulin Misfolding Is an Early Event in the Progression to Type 2 Diabetes," *eLife* 8 (2019): e44532, https://doi.org/10.7554/eLife.44532.
[54] P. Sweeney et al., "Protein misfolding in neurodegenerative diseases: implications and strategies," *Translational Neurodegeneration* 6 (2017): 6, https://doi.org/10.1186/s40035-017-0077-5.

Gaining mechanistic insight into folding and misfolding kinetics may aid in differentiating pathogenic trajectories from stable molecular alterations.

In this article, we revisit the conceptual foundations of protein folding mechanisms and situate them within a broader framework of conformational dynamics and macromolecular self-assembly. We critically examine emerging experimental methodologies that interrogate folding and misfolding across spatial and temporal scales, and review recent advances in computational modeling and artificial intelligence (AI), highlighting both achievements and persistent limitations. We identify the key gaps that that continue to constrain the development of a predictive theory of folding dynamics and propose a roadmap for integrating experimental and computational strategies into a unified framework. Finally, we articulate the scientific and translational implications of achieving a quantitative, predictive understanding of protein folding for biology, biotechnology, and medicine.

# 1. The Paradox of Structure Prediction Without Folding Mechanism

Although current machine learning methods, such as AlphaFold and RoseTTAFold, provide accurate structural predictions for a large fraction of the proteome, such structure prediction resolves only one dimension of the mutifaceted protein folding issue. While structure prediction provides a model of the native state, it does not explain how that state is reached, how alternative states are accessed, or how folding proceeds in time and in the cellular environment. The central paradox is therefore that we can often predict one or few of the functional states of a protein, but we cannot predict the complete structural ensemble, the folding pathways and their kinetics, how a protein interconverts between functional states and assembles into complexes, or protein susceptibility to misfolding (Fig. 1). Resolving this gap requires moving from static structure determination to a quantitative description of folding as a dynamic, stochastic, and context-dependent process.

## 1.1 The recent revolution in structure prediction has transformed structural biology

The 3D structure prediction of many proteins directly from their amino acid sequence rests on two distinct pillars: the systematic accumulation of experimentally-determined protein structures, which created a dense and diverse structural repository, and advances in neural network architectures capable of extracting high-dimensional correlations from large multiple sequence alignments. The Protein Data Bank[55] provided the essential training corpus. X-ray crystallography,[56] NMR spectroscopy,[57] and cryo-EM[58] generated structural coverage across major protein families, enabling machine learning models to infer statistical constraints linking sequence evolution to spatial organization. The fundamental information that is used in these methods largely comes from extracting the information from these multiple-sequence alignments, using the idea that residue pairs that are in contact tend to co-evolve[59] and the fact that structure is in general conserved throughout evolution.[60]

[55] H. M. Berman, "The Protein Data Bank," *Nucleic Acids Research* 28 (January 2000): 235–42, https://doi.org/10.1093/nar/28.1.235.
[56] A. Ilari and C. Savino, "Protein Structure Determination by X-Ray Crystallography," in *Bioinformatics: Data, Sequence Analysis and Evolution* (2008), https://doi.org/10.1007/978-1-60327-159-2_3.
[57] K. Wüthrich, "The Way to NMR Structures of Proteins," *Nature Structural Biology* 8 (2001): 923–25, https://doi.org/10.1038/nsb1101-923.
[58] W. Kühlbrandt, "The resolution revolution," *Science* 343 (2014): 1443–44, https://doi.org/10.1126/science.1251652.

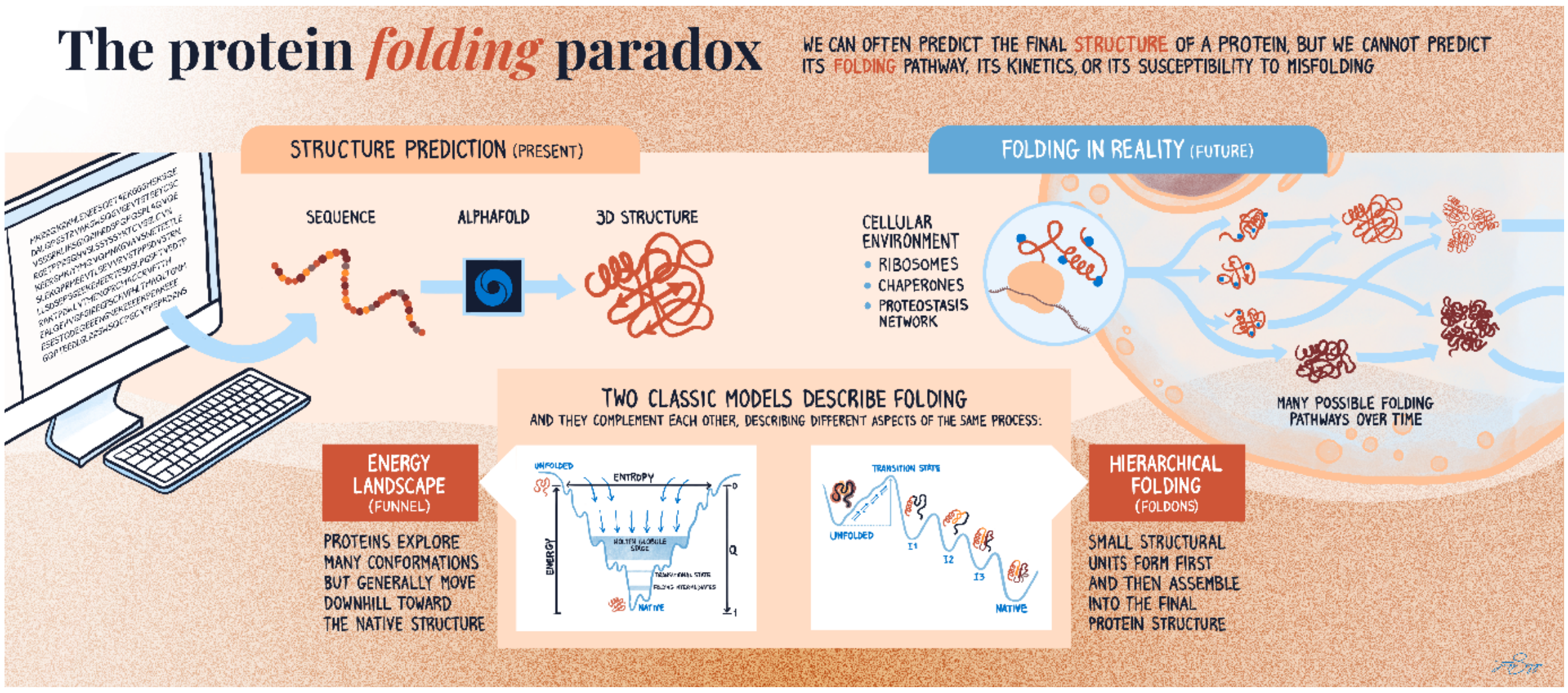

*Figure 1.* ***The protein folding paradox****. Currently, models exist that describe protein folding. However, despite the advances in experimental methods and AI towards revealing protein structure, we are still lacking a detailed understanding of dynamics and precise folding pathways. Graphics by J. Saquegno.*

The impact new computational methods to predict protein structure has been immediate: accurate structural models can now be produced at proteome scale with unprecedented speed.[61] Structural annotation has become feasible for entire genomes; structure-guided drug discovery[62] and recombinant protein design[63] demonstrate improved success rates in situations that lack prior experimental structures.[64] Despite these advances, current models remain fundamentally centered on predicting structural states rather than fully characterizing the underlying thermodynamic and kinetic landscapes. Emerging approaches are beginning to move beyond single static structures toward the prediction of conformational ensembles, but their accuracy and generality remain difficult to assess. The central challenge is therefore shifting from predicting structures to validating at scale the thermodynamics and kinetics of protein conformational landscapes. Addressing this challenge will require systematic integration of high-resolution experimental data with computational models, enabling rigorous benchmarking of both equilibrium ensembles and dynamic pathways.

---

[59] U. Göbel et al., "Correlated Mutations and Residue Contacts in Proteins," *Proteins: Structure, Function, and Bioinformatics* 18 (1994): 309–17, https://doi.org/10.1002/prot.340180402.
[60] K. Illergård et al., "Structure Is Three to Ten Times More Conserved than Sequence—a Study of Structural Response in Protein Cores," *Proteins: Structure, Function, and Bioinformatics* 77 (2009): 499–508, https://doi.org/10.1002/prot.22458.
[61] O. Kovalevskiy et al., "AlphaFold Two Years on: Validation and Impact," *Proceedings of the National Academy of Sciences* 121 (2024): e2315002121, https://doi.org/10.1073/pnas.2315002121.
[62] A. Mullard, "What Does AlphaFold Mean for Drug Discovery?," *Nature Reviews Drug Discovery* 20 (2021): 725–28, https://doi.org/10.1038/d41573-021-00161-0.
[63] J. L. Watson et al., "De Novo Design of Protein Structure and Function with RFdiffusion," *Nature* 620 (2023): 1089–100, https://doi.org/10.1038/s41586-023-06415-8.
[64] F. Ren et al., "AlphaFold Accelerates Artificial Intelligence Powered Drug Discovery: Efficient Discovery of a Novel CDK20 Small Molecule Inhibitor," *Chemical Science* 14 (2023): 1443–52, https://doi.org/10.1039/D2SC05709C.

## 1.2 Predicting static structures does not explain folding pathways or kinetics

Protein folding is a time-dependent process governed by an underlying energy landscape that encodes multiple possible routes from unfolded states to the native structure. Along these trajectories, a polypeptide chain may populate partially folded intermediates, transient transition states, and off-pathway conformations. The kinetics of barrier crossing and the topology of the landscape determine folding rates, pathway heterogeneity, and the likelihood of misfolding.

Static structure prediction provides no direct information about these dynamical features. It does not specify the order in which structural elements form, the identities and lifetimes of intermediates, or the magnitude of free-energy barriers separating metastable states. Nor does it predict folding and unfolding rate constants or quantify the probability of aggregation under given conditions.

*In vivo*, additional layers of regulation further complicate the process. Folding frequently begins co-translationally as the nascent chain emerges from the ribosome, imposing a sequential and vectorial constraint on structural formation.[65] Ribosome-associated factors and molecular chaperones interact with partly synthesized chains, modulating conformational sampling and suppressing formation of misfolded protein aggregates, while still enabling multimerization of individual protein molecules into oligomeric states that define protein function.[66] These interactions can alter folding pathways, delay compaction, or stabilize specific intermediates.[67] Critically, while protein aggregation is a hallmark of various pathological states, controlled formation of higher stoichiometry protein clusters may be physiological with a purpose of cellular adaptation to the environment,[68] gene regulation,[69] storing peptides and hormones, cell cycle regulation, microbial adhesion, biofilm formation, and host invasion.[70] [71] [72] [73]

---

[65] C. A. Waudby et al., “Nature and Regulation of Protein Folding on the Ribosome,” *Trends in Biochemical Sciences* 44 (2019): 914–26, https://doi.org/10.1016/j.tibs.2019.06.008.
[66] F. U. Hartl et al., “Molecular Chaperones in Protein Folding and Proteostasis,” *Nature* 475 (2011): 324–32, https://doi.org/10.1038/nature10317.
[67] S. H. S. Chan et al., “The Ribosome Stabilizes Partially Folded Intermediates of a Nascent Multi-Domain Protein,” *Nature Chemistry* 14 (2022): 1165–73, https://doi.org/10.1038/s41557-022-01004-0.
[68] S. Shashkova et al., “Correlative Single-Molecule Fluorescence Barcoding of Gene Regulation in Saccharomyces Cerevisiae,” *Methods* 193 (2021): 62–67, https://doi.org/10.1016/j.ymeth.2020.10.009.
[69] Adam Jm Wollman et al., “Transcription Factor Clusters Regulate Genes in Eukaryotic Cells,” *eLife* 6 (August 2017): e27451, https://doi.org/10.7554/eLife.27451.
[70] S. K. Maji et al., “Functional Amyloids as Natural Storage of Peptide Hormones in Pituitary Secretory Granules,” *Science* 325 (2009): 328–32, https://doi.org/10.1126/science.1173155.
[71] S. Saad et al., “Reversible Protein Aggregation Is a Protective Mechanism to Ensure Cell Cycle Restart after Stress,” *Nature Cell Biology* 19 (2017): 1202–13, https://doi.org/10.1038/ncb3600.
[72] D. M. Fowler et al., “Functional Amyloid–from Bacteria to Humans,” *Trends in Biochemical Sciences* 32 (2007): 217–24, https://doi.org/10.1016/j.tibs.2007.03.003.
[73] Martijn F. B. G. Gebbink et al., “Amyloids — a Functional Coat for Microorganisms,” *Nature Reviews Microbiology* 3 (April 2005): 333–41, https://doi.org/10.1038/nrmicro1127.

Misfolding and aggregation provide a further illustration of the limitation of endpoint prediction. Proteins associated with neurodegenerative diseases often adopt alternative conformations that nucleate oligomerization or fibril formation.[23] These processes are governed by kinetic competition between physiological folding and pathological aggregation pathways.[74] A correct prediction of the native structure does not determine whether a protein will fold efficiently in the cellular environment or whether it will access pathogenic states.

### 1.3 Proteins function as dynamic ensembles rather than single static structures

Even after folding, proteins do not exist as rigid entities. Native states are characterized by an ensemble of conformations that span multiple states separated by free-energy differences.[75] Functional processes such as ligand binding, catalysis, and allosteric regulation often involve shifts within this ensemble rather than transitions between fully unfolded and fully folded states.[76] A single structural model captures only one representative configuration within a broader probability distribution. Efforts towards large-scale emulation of protein conformational dynamic characteristics are emerging but are coarse-grained and do not resolve time-dependent evolution, nor capture all the relevant meta-stable conformations.[77] More recent emulators target also time-dependent statistics but are coarse-grained and do not scale to larger systems.[78]

Intrinsically disordered proteins further emphasize this point. Many proteins or domains lack a uniquely defined 3D structure under physiological conditions.[79] Instead, they populate heterogeneous ensembles stabilized by transient intramolecular interactions or by binding partners.[80] The structural heterogeneity of such regions is reflected in the lower confidence of prediction algorithms, but this variability is a biological property rather than a failure of modeling.

Metastable conformations and alternative assemblies also play central roles in disease. The same amino acid sequence can give rise to multiple structural polymorphs, particularly in amyloid-forming proteins, where distinct fibrillar architectures are associated with different pathological phenotypes.[81] In the Amyloid Atlas database[82] and related website,[83] there are currently 800 entries whereof 86 are Aβ, 202 are Tau, and 191 are α-synuclein high resolution amyloid fibril structures. Most fibrils have divergent folds, which clearly demonstrates that neurodegenerative proteins show vastly polymorphic aggregation behaviors: today, these are entirely unpredictable from static models. Even highly stable globular proteins such as IgG light chains and transthyretin can completely refold and form the misfolded amyloid fibril state. These examples illustrate the limitation of interpreting Anfinsen's hypothesis as implying a single biologically relevant endpoint under all conditions. With current knowledge, it cannot be understood how this dramatic conformational conversion occurs, which appears to entail complete unfolding of the native protein structure prior to misfolding into the amyloid fibril state.[84,85,86]

[74] Z. Jia et al., "Amyloid Assembly Is Dominated by Misregistered Kinetic Traps on an Unbiased Energy Landscape," *Proceedings of the National Academy of Sciences* 117 (2020): 10322–28, https://doi.org/10.1073/pnas.1911153117.
[75] R. Nussinov et al., "Protein Conformational Ensembles in Function: Roles and Mechanisms," *RSC Chemical Biology* 4 (2023): 850–64, https://doi.org/10.1039/D3CB00114H.
[76] H. Motlagh et al., "The Ensemble Nature of Allostery," *Nature* 508 (2014): 331–39, https://doi.org/10.1038/nature13001.
[77] S. Lewis et al., "Scalable Emulation of Protein Equilibrium Ensembles with Generative Deep Learning," *Science* 389 (2025): eadv9817, https://doi.org/10.1126/science.adv9817.
[78] P. Antoniadis et al., "Protein Language Model Embeddings Improve Generalization of Implicit Transfer Operators," ICML, 2026.
[79] A. K. Dunker et al., "Intrinsically Disordered Protein," *Journal of Molecular Graphics and Modelling* 19 (2001): 26–59, https://doi.org/10.1016/S1093-3263(00)00138-8.
[80] A. S. Holehouse and B. B. Kragelund, "The Molecular Basis for Cellular Function of Intrinsically Disordered Protein Regions," *Nature Reviews Molecular Cell Biology* 25 (2024): 187–211, https://doi.org/10.1038/s41580-023-00673-0.

A comprehensive understanding of protein behavior therefore requires describing conformational distributions, the transitions between them, and the interactions that give rise to higher-order assemblies. In living cells, the distribution in the context of organelle localization indicates the functional role of a given protein distribution pattern. For example, which amyloid fibril localizes to a given mitochondrial respiration complex and disrupts its activity, or which organelle contact sites are implicated, will reveal major insights into cell function and pathogenesis.[87] Hence, the relevant object is not an individual structure, but a statistical ensemble evolving over time and across interacting molecular partners. Predicting these ensembles, their kinetics, and their assembly into functional complexes from sequence remains an open challenge.

# 2. Folding as a Multiscale and Context-Dependent Process

Protein folding is not an isolated thermodynamic event, but a multiscale process embedded within the cellular environment. The physical interactions that drive local secondary structure formation operate at the Ångström and nanosecond scale, whereas folding outcomes influence cellular function over minutes, years, or even decades. Experimental and computational approaches typically capture only limited portions of this hierarchy. Bridging these scales remains one of the central challenges in developing a predictive theory of folding. Therefore, folding must be understood simultaneously as a molecular, cellular, and organismal phenomenon as the mechanistic determinants of folding pathways cannot be separated from the context in which folding occurs (Fig. 2).

[81] L. D. Aubrey and S. E. Radford, "How Is the Amyloid Fold Built? Polymorphism and the Microscopic Mechanisms of Fibril Assembly," *Journal of Molecular Biology* 437 (2025): 169008, https://doi.org/10.1016/j.jmb.2025.169008.
[82] M. R. Sawaya et al., "The Expanding Amyloid Family: Structure, Stability, Function, and Pathogenesis," *Cell* 184 (2021): 4857–73, https://doi.org/10.1016/j.cell.2021.08.013.
[83] Sawaya Group, "Amyloid Atlas," UCLA, https://people.mbi.ucla.edu/sawaya/amyloidatlas/.
[84] L. Radamaker et al., "Cryo-EM Structure of a Light Chain-Derived Amyloid Fibril from a Patient with Systemic AL Amyloidosis," *Nature Communications* 10 (2019): 1103, https://doi.org/10.1038/s41467-019-09032-0.
[85] P. Swuec et al., "Cryo-EM structure of cardiac amyloid fibrils from an immunoglobulin light chain AL amyloidosis patient," *Nature Communications* 10 (2019): 1269, https://doi.org/10.1038/s41467-019-09133-w.
[86] M. Schmidt et al., "Cryo-EM Structure of a Transthyretin-Derived Amyloid Fibril from a Patient with Hereditary ATTR Amyloidosis," *Nature Communications* 10 (2019): 5008, https://doi.org/10.1038/s41467-019-13038-z.
[87] Lucia Pagani and Anne Eckert, "Amyloid-Beta Interaction with Mitochondria," *International Journal of Alzheimer's Disease* 12 (2011): 925050, https://doi.org/10.4061/2011/925050.

## 2.1 Folding spans spatial and temporal scales that remain difficult to integrate

At the smallest scale, folding is governed by atomic-level interactions among amino acid side chains, backbone hydrogen bonds, electrostatics, and solvent-mediated effects. These interactions fluctuate on femtosecond-to-nanosecond timescales and define the local energy landscape explored by the polypeptide chain.[88] Molecular dynamics simulations have successfully captured aspects of this regime for small proteins, but the computational cost increases steeply with system size and timescale, limiting routine access to folding events for larger or slower-folding proteins.[89]

At intermediate scales, folding involves the cooperative assembly of structural elements such as α-helices and β-sheets, domain formation, and tertiary packing. These processes occur over microseconds to seconds and often involve partly structured intermediates.[90] Experimental techniques—such as turbulent flow mixers in stopped-flow or microfluidics devices,[91] [92] HDX,[93] and single-molecule force spectroscopy[94]—provide access to portions of this regime, but each approach samples different observables and resolutions. Integrating these measurements into a unified kinetic model remains nontrivial.

At larger scales, protein folding occurs within distinct cellular environments including the crowded cytoplasm, organelles, membrane interfaces, vesicular compartments, and along the secretory pathway. At the same time, the extracellular microenvironment as well as cellular compartments impose additional constraints on molecular concentrations, redox conditions, pH, ion composition, cytoplasmic viscosity, and molecular interactions. These environmental factors reshape folding kinetics, conformational stability, trafficking, degradation, and aggregation behavior. Because proteins operate within interconnected signaling networks, misfolding and aggregation can propagate beyond the originating cell through secretion, vesicular trafficking, extracellular vesicles, or direct cell-to-cell transfer, ultimately affecting tissue-level function.[95]

[88] B. Fierz et al., "Loop Formation in Unfolded Polypeptide Chains on the Picoseconds to Microseconds Time Scale," *Proceedings of the National Academy of Sciences* 104 (2007): 2163-2168, https://doi.org/10.1073/pnas.0611087104.
[89] J. Viguera Diez et al., "Transferable Generative Models Bridge Femtosecond to Nanosecond Time-Step Molecular Dynamics," *Science Advances* 12 (2026): eaed2333, https://doi.org/10.1126/sciadv.aed2333.
[90] A. V. Glyakina and O. V. Galzitskaya, "How Quickly Do Proteins Fold and Unfold, and What Structural Parameters Correlate with These Values?," *Biomolecules* 10 (2020): 197, https://doi.org/10.3390/biom10020197.
[91] V. Inguva et al., "Computer Design of Microfluidic Mixers for Protein/RNA Folding Studies," *PLOS ONE* 13 (2018): e0198534, https://doi.org/10.1371/journal.pone.0198534.
[92] S. A. Waldauer et al., "Microfluidic Mixers for Studying Protein Folding," *Journal of Visualized Experiments* 62 (2012): e3976, https://doi.org/10.3791/3976.
[93] R. Li and C. Woodward, "The Hydrogen Exchange Core and Protein Folding," *Protein Science* 8 (1999): 1571–90, https://doi.org/10.1110/ps.8.8.1571.
[94] R. Petrosyan et al., "Single-molecule force spectroscopy of protein folding," *Journal of Molecular Biology* 433 (2021): 167207, https://doi.org/10.1016/j.jmb.2021.167207.

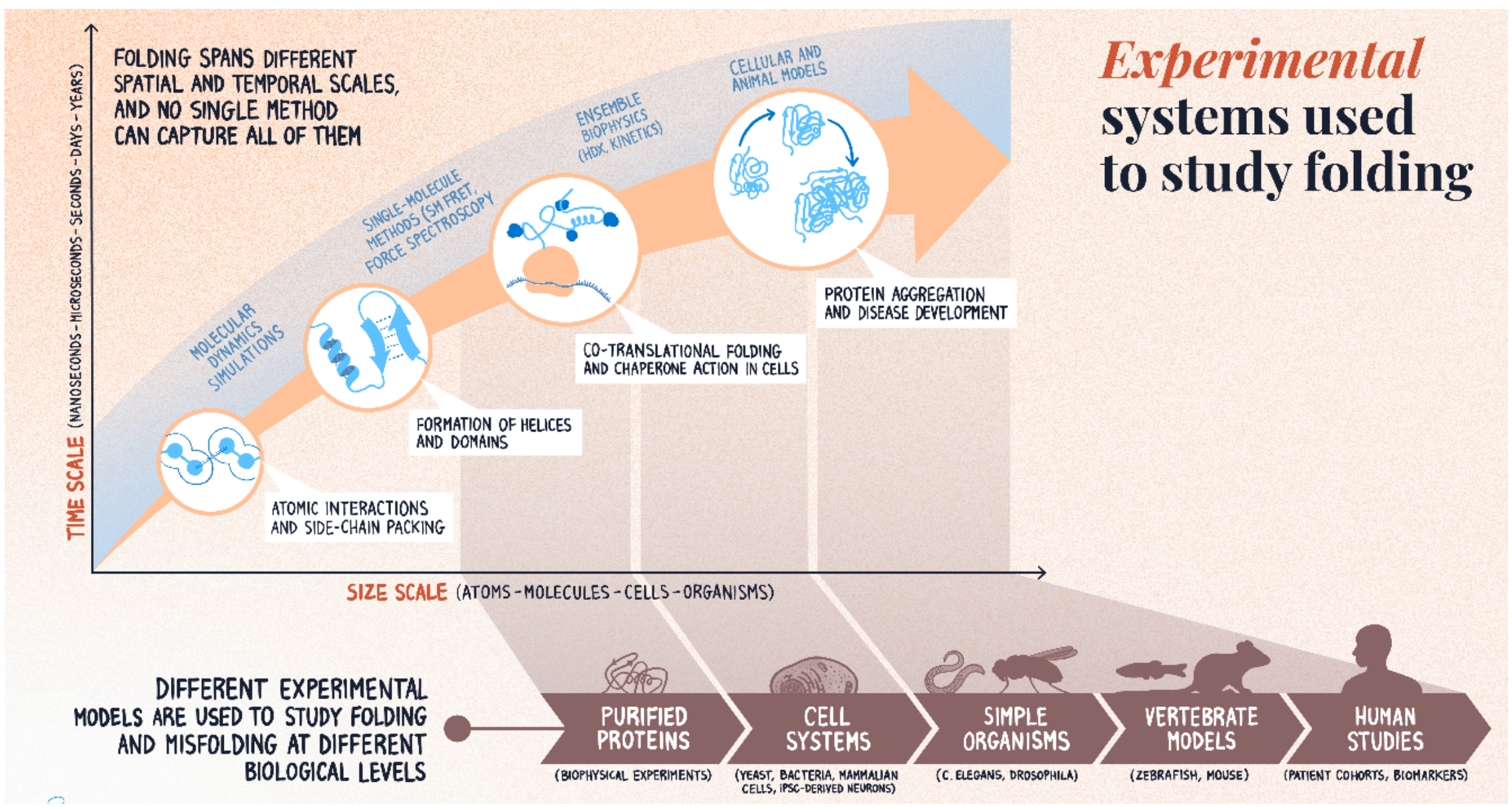

*Figure 2.* ***Experimental systems in folding studies****. Mechanistic determinants of folding pathways cannot be separated from the context in which folding occurs. Hence, folding must be understood simultaneously at a molecular, cellular, and organismal levels. However, bridging time and size scales remains one of the central challenges in developing a predictive theory of folding. Graphics by J. Saquegno.*

Temporal scales pose an equally significant challenge. Elementary folding transitions may occur on microsecond timescales, yet disease-related misfolding processes develop over years or decades. Neurodegenerative disorders such as Alzheimer's and Parkinson's diseases are associated with the gradual accumulation of misfolded aggregates, often long before clinical symptoms appear.[96] Experimental systems necessarily compress these timescales. Yeast, fruit fly, zebrafish, and mouse models provide valuable insights, but they operate over hours, days, or months, requiring extrapolation to human lifespans.

## 2.2 Folding in cells is shaped by translation, molecular chaperones, and proteostasis networks

*In vivo*, folding frequently begins before synthesis is complete. As the nascent polypeptide chain emerges from the ribosome, it experiences vectorial growth and spatial confinement.[97,98] This sequential exposure of residues imposes kinetic ordering on structure formation that differs from refolding experiments initiated from fully denatured states *in vitro*.[99] The ribosome itself can influence conformational sampling through steric constraints and interactions with the exit tunnel.[97] In addition, codon usage influences translation speed.[100]

---

[95] N. Brezic et al., "Protein Misfolding and Aggregation as a Mechanistic Link Between Chronic Pain and Neurodegenerative Diseases," *Current Issues in Molecular Biology* 47 (2025): 259, https://doi.org/10.3390/cimb47040259.
[96] W. Fu and P. C. L. Ho, "Blood-Based Biomarkers for Alzheimer's Disease: Advances in Early Detection and Monitoring of Age-Related Neurodegeneration," *Ageing Research Reviews*, 2026, 103058, https://doi.org/10.1016/j.arr.2026.103058.
[97] M. Liutkute et al., "Cotranslational Folding of Proteins on the Ribosome," *Biomolecules* 10 (2020): 97, https://doi.org/10.3390/biom10010097.
[98] S. Wang et al., "Cotranslational Protein Folding through Non-Native Structural Intermediates," *Science Advances* 11 (2025): eady2211, https://doi.org/10.1126/sciadv.ady2211.

Molecular chaperones further modulate folding pathways. Ribosome-associated factors interact early during the synthesis with nascent chains[101] and chaperones, such as trigger factor, NAC (Nascent Chain-Associated Complex), Hsp70/40 (Heat Shock Protein 70 and 40 Complex), Hsp90 (Heat Shock Protein 90), and chaperonins, bind exposed hydrophobic regions and prevent premature aggregation.[102] Some polypeptide chains require iterative cycles of binding and release to reach their native conformation. Others are directed toward degradation if folding fails. Chaperones also facilitate posttranslational modifications (PTMs), such as phosphorylation and glycosylation, which further shape proteins and define their function, including subsequent aggregation and disease pathology.[103] Some PTMs may occur co-translationally. These systems do not simply stabilize the final structure; they also alter kinetic partitioning among competing pathways.

Proteostasis networks integrate translation, folding, trafficking, and degradation. The balance among synthesis, folding efficiency, chaperone capacity, and clearance mechanisms determines the steady-state distribution of protein conformations within the cell.[104] Perturbations of this equilibrium, whether due to mutation, stress, or aging, can promote the misfolded or aggregated states.[105,106] It should be noted also that misfolding during aging may be due to mis-translation of newly translated proteins rather than mutations, or to old, long-lived proteins starting to unfold. This may be an important contributor to aging, as studies have shown that mis-translation is increased upon aging and that mutations or drugs that boost ribosomal proofreading, for example in mitochondria, can prolong lifespan,[107] and that translation-error-prone mice mutants accumulate aggregates faster and age prematurely.[108]

---

[99] K. Liu et al., “The Ribosome Destabilizes Native and Non-Native Structures in a Nascent Multidomain Protein,” *Protein Science* 26 (2017): 1439–51, https://doi.org/10.1002/pro.3189.
[100] F. Buhr et al., “Synonymous Codons Direct Cotranslational Folding toward Different Protein Conformations,” *Molecular Cell* 61 (2016): 341–51, https://doi.org/10.1016/j.molcel.2016.01.008.
[101] M. Gamerdinger et al., “Ribosome-NAC Collaboration: A Regulatory Platform for Cotranslational Chaperones, Enzymes, and Targeting Factors,” *Molecular Cell* 86 (2026): 491–502, https://doi.org/10.1016/j.molcel.2025.12.031.
[102] D. Balchin et al., “In Vivo Aspects of Protein Folding and Quality Control,” *Science* 353 (2016): aac4354, https://doi.org/10.1126/science.aac4354.
[103] C. Alquezar et al., “Tau Post-Translational Modifications: Dynamic Transformers of Tau Function, Degradation, and Aggregation,” *Frontiers in Neurology* 11 (2021): 595532, https://doi.org/10.3389/fneur.2020.595532.
[104] W. E. Balch et al., “Adapting Proteostasis for Disease Intervention,” *Science* 319 (2008): 916–19, https://doi.org/10.1126/science.1141448.
[105] S. Dukan et al., “Protein Oxidation in Response to Increased Transcriptional or Translational Errors,” *Proceedings of the National Academy of Sciences* 97 (2000): 5746–49, https://doi.org/10.1073/pnas.100422497.
[106] V. Kohler and C. Andréasson, “Reversible Protein Assemblies in the Proteostasis Network in Health and Disease,” *Frontiers in Molecular Biosciences* 10 (2023): 1155521, https://doi.org/10.3389/fmolb.2023.1155521.
[107] T. Suhm et al., “Mitochondrial Translation Efficiency Controls Cytoplasmic Protein Homeostasis,” *Cell Metabolism* 27 (2018): 1309–22, https://doi.org/10.1016/j.cmet.2018.04.011.
[108] Erik C. Böttger et al., “Translational Error in Mice Increases with Ageing in an Organ-Dependent Manner,” *Nature Communications* 16 (February 2025): 2069, https://doi.org/10.1038/s41467-025-57203-z.

Environmental and biochemical stressors can further reshape folding landscapes and protein homeostasis, such as alterations in metal ion homeostasis, redox balance, temperature, or pH.[109] Typically, the cellular protein quality control machinery is capable of resolving these aggregates; however, their accumulation often leads to disease development and is serving as an aging hallmark.

Folding cannot be decoupled from this regulatory environment. Different proteins follow distinct cellular handling pathways. The fate of a misfolding-prone protein depends not only on its intrinsic energy landscape, but also on cell type, stress conditions, and the state of the protein quality control machinery.[109] Consequently, folding mechanisms inferred from dilute *in vitro* experiments may not fully capture the dynamics operating in living systems. Understanding folding in context therefore requires coupling molecular-level measurements with cellular models and computational representations that incorporate translation kinetics, chaperone interactions, and degradation pathways.

### 2.3 Misfolding pathways are central to disease and cannot be inferred from structure alone

Misfolding and aggregation are not rare anomalies, but central determinants of numerous diseases.[110] In neurodegenerative disorders, specific proteins adopt alternative conformations that nucleate oligomerization and fibril formation.[111] Some of these assemblies can propagate through templated conformational conversion, giving rise to prion-like spreading across tissues.[112] The same amino acid sequence can assemble into multiple fibrillar architectures, each associated with distinct clinical manifestations.[113] Such diversity arises from the underlying folding landscape and kinetic accessibility of alternative states. The pathogenic process depends on kinetic competition between native folding and off-pathway aggregation. A protein that is structurally well-defined in its native state may nonetheless be prone to misfolding under stress conditions.[114] The static native structure does not reveal the height of aggregation barriers, the stability of partially folded intermediates, or the existence of low-population conformers that might seed aggregation.

Degradation of aggregated protein is often controlled by the process of autophagy, a critical intracellular protein degradation pathway for long-lived proteins. With most of the cell's proteins being long-lived, autophagy is not only controlling intracellular proteostasis but it is also positioned as a cellular stress response that enables the supply of amino acids and metabolite substrates during metabolic perturbation or nutrient deprivation. Moreover, it is functionally crucial for the removal of dysfunctional mitochondria, playing a critical role in mitochondrial quality control. Neurons are particularly sensitive to changes in protein homeostasis, due to their high reliance on protein synthesis and degradation and terminally differentiated nature, making them extremely vulnerable to toxic protein accumulation.[115] Autophagy failure and the aggregation of protein aggregates constitute a core hallmark of brain aging,[115] with aging being the biggest risk factor for the onset of neurodegenerative disease, including Alzheimer's disease and Parkinson's disease. It is estimated that by 2050, the global population of the age above 60 will reach 22%, with the majority living in developing countries. This places the global South at particular risk, with poor aging being exacerbated due to respective socioeconomic vulnerabilities.[116] The improvement in longevity has also heightened the susceptibility to age-associated pathologies such as neurodegeneration.[117,118] Particularly in Africa, neurodegenerative diseases impact strongly the function of communities, due to the stigma associated with brain illness.[119]

---

[109] Fabrizio Chiti and Christopher M. Dobson, "Protein Misfolding, Amyloid Formation, and Human Disease: A Summary of Progress Over the Last Decade," *Annual Review of Biochemistry* 86 (2017): 27–68, https://doi.org/10.1146/annurev-biochem-061516-045115.
[110] M. R. Ajmal, "Protein Misfolding and Aggregation in Proteinopathies: Causes, Mechanism and Cellular Response," *Diseases* 11 (2023): 30, https://doi.org/10.3390/diseases11010030.
[111] O. Koszła and P. Sołek, "Misfolding and Aggregation in Neurodegenerative Diseases: Protein Quality Control Machinery as Potential Therapeutic Clearance Pathways," *Cell Communication and Signaling* 22 (2024): 421, https://doi.org/10.1186/s12964-024-01791-8.
[112] S. K. Kaufman and M. I. Diamond, "Prion-like Propagation of Protein Aggregation and Related Therapeutic Strategies," *Neurotherapeutics* 10 (2013): 371–82, https://doi.org/10.1007/s13311-013-0196-3.
[113] M. Fändrich et al., "Amyloid Fibril Polymorphism: A Challenge for Molecular Imaging and Therapy," *Journal of Internal Medicine* 283 (March 2018): 218–37, https://doi.org/10.1111/joim.12732.
[114] N. Louros et al., "Mechanisms and Pathology of Protein Misfolding and Aggregation," *Nature Reviews Molecular Cell Biology* 24 (2023): 912–33, https://doi.org/10.1038/s41580-023-00647-2.

Animal and cellular models such as yeast, mammalian cells, fruit flies, zebrafish, and mice, provide evidence that aggregation pathways are sensitive to expression levels, intracellular condensation (in terms of local concentrations), cellular environment, and clearance capacity.[110,120,121] Yeast, in particular, has proven instrumental in elucidating prion-like behavior and proteostasis stress responses on a large scale.[122] Meanwhile, zebrafish and mouse models facilitate *in vivo* imaging, tissue-specific manipulation and the investigation of disease-relevant phenotypes in complex organisms.[123] However, the temporal resolution achievable *in vivo* often limits direct observation of protein kinetics and dynamics.[124] Mutational studies can probe the contribution of specific residues in folding and unfolding, but they report on ensemble behavior, rather than single-molecule trajectories.[125] Single-molecule microscopy studies are able to determine molecular subpopulations; however, typically such approaches have limited throughput.[121] Bridging these gaps requires integrating molecular-resolution measurements and high throughput methods with cellular- and organism-level phenotypes.

[115] B. Loos et al., "Augmenting Brain Metabolism to Increase Macro-and Chaperone-Mediated Autophagy for Decreasing Neuronal Proteotoxicity and Aging," *Progress in Neurobiology* 156 (2017): 90–106, https://doi.org/10.1016/j.pneurobio.2017.05.001.
[116] C. Ntsapi and B. Loos, "Caloric Restriction and the Precision-Control of Autophagy: A Strategy for Delaying Neurodegenerative Disease Progression," *Experimental Gerontology* 83 (2016): 97–111, https://doi.org/10.1016/j.exger.2016.07.014.
[117] J. S. Bhatti et al., "Therapeutic Strategies for Mitochondrial Dysfunction and Oxidative Stress in Age-Related Metabolic Disorders," *Progress in Molecular Biology and Translational Science* 146 (2017): 13–46, https://doi.org/10.1016/bs.pmbts.2016.12.012.
[118] T. Maduna et al., "Macroautophagy and Chaperone-Mediated Autophagy in Aging," in *Factors Affecting Neurological Aging* (Academic Press, 2021), https://doi.org/10.1016/B978-0-12-817990-1.00018-4.
[119] T. Maduna and B. Loos, "Alzheimer's Disease—Molecular Defect, Public Perceptions and Stigma in South Africa," in *Health Communication and Disease in Africa*, ed. B. Falade and M. Murire (Palgrave Macmillan, 2021), https://doi.org/10.1007/978-981-16-2546-6_4.
[120] R. Li et al., "Single-Molecule Dynamics of the TRiC Chaperonin System in Vivo," *Nature* 652 (2026): 481–89, https://doi.org/10.1038/s41586-025-10073-3.
[121] M. Mao et al., "Controlling Protein Stability with SULI, a Highly Sensitive Tag for Stabilization upon Light Induction," *Nature Communications* 14 (2023): 2172, https://doi.org/10.1038/s41467-023-37830-0.
[122] C. Batlle et al., "Characterization of Soft Amyloid Cores in Human Prion-Like Proteins," *Scientific Reports* 7 (2017): 12134, https://doi.org/10.1038/s41598-017-09714-z.
[123] D. Imberechts et al., "Established and Emerging New Approach Methodologies in Neuroscience," *Frontiers in Neuroscience* 19 (2025): 1696937, https://doi.org/10.3389/fnins.2025.1696937.
[124] Takahiro Deguchi et al., "Direct Observation of Motor Protein Stepping in Living Cells Using MINFLUX," *Science* 379 (March 2023): 1010–15, https://doi.org/10.1126/science.ade2676.

A mechanistic understanding of misfolding therefore demands explicit characterization of folding trajectories, intermediate populations, and aggregation kinetics. Disease phenotypes emerge from dynamic processes operating over extended timescales. Without quantitative insights into these dynamics, structure-based predictions remain incomplete. Clinical biomarker studies further illustrate this limitation. Blood-based biomarkers for Alzheimer's disease, such as phosphorylated tau isoforms,[126] neurofilament light chain,[127] and GFAP,[128] correlate with brain pathology and predict progression in selected specialist cohorts. However, their predictive value decreases in community-based populations, particularly in asymptomatic individuals.[129] Comorbidities, renal function, and body mass index substantially influence circulating protein levels.[130,131] In addition, assay performance can vary as different methods are prone to detect different protein variants.[132] As a result, biomarker thresholds are context-dependent and often insufficient for robust staging outside highly controlled settings.[133] These observations highlight the need for more mechanistically grounded markers that reflect specific conformational states or kinetic vulnerabilities rather than aggregate protein abundance alone.

---

[125] K. L. Schneider et al., "Studying Spatial Protein Quality Control, Proteopathies, and Aging Using Different Model Misfolding Proteins in S. Cerevisiae," *Frontiers in Molecular Neuroscience* 11 (2018): 249, https://doi.org/10.3389/fnmol.2018.00249.
[126] H. Hampel et al., "Blood-Based Biomarkers for Alzheimer's Disease: Current State and Future Use in a Transformed Global Healthcare Landscape," *Neuron* 111 (2023): 2781–99, https://doi.org/10.1016/j.neuron.2023.05.017.
[127] E. Civita et al., "Advancing Clinical Use of Neurofilament Light Chain: Translational Insights from Research to Routine Practice," *Biomarker Insights* 20 (2025): 11772719251364018, https://doi.org/10.1177/11772719251364018.
[128] E. M. S. Bandara et al., "The Role of Glial Fibrillary Acidic Protein in the Neuropathology of Alzheimer's Disease and Its Potential as a Blood Biomarker for Early Diagnosis and Progression," *Molecular Neurobiology* 62 (2025): 15576–608, https://doi.org/10.1007/s12035-025-05219-3.
[129] B. Khorsand et al., "Incremental Value of Plasma Biomarkers in Predicting Clinical Decline among Cognitively Unimpaired Older Adults: Results from the A4 Trial," *Alzheimer's & Dementia* 18 (2026): e70321, https://doi.org/10.1002/dad2.70321.
[130] X. Gong et al., "The Impact of Weight Loss on Renal Function in Individuals with Obesity and Type 2 Diabetes: A Comprehensive Review," *Frontiers in Endocrinology* 15 (2024): 1320627, https://doi.org/10.3389/fendo.2024.1320627.
[131] J. F. Wan et al., "Impact of Body Mass Index on Adverse Kidney Events in Diabetes Mellitus Patients: A Systematic-Review and Meta-Analysis," *World Journal of Clinical Cases* 12 (2024): 538–50, https://doi.org/10.12998/wjcc.v12.i3.538.
[132] D. Gogishvili et al., "The GFAP Proteoform Puzzle: How to Advance GFAP as a Fluid Biomarker in Neurological Diseases," *Journal of Neurochemistry* 169 (2025): 16226, https://doi.org/10.1111/jnc.16226.
[133] Q. Zhao et al., "Applications and Challenges of Biomarker-Based Predictive Models in Proactive Health Management," *Frontiers in Public Health* 13 (2025): 1633487, https://doi.org/10.3389/fpubh.2025.1633487.

# 3. Experimental Approaches Are Beginning to Resolve Folding Trajectories

A major obstacle to predictive models of protein folding has been the limited availability of quantitative, time-resolved folding trajectory data. To date, folding dynamics have mostly been characterized through ensemble observables or indirect proxies. As a result, kinetic models often rely on simplified state representations and reduced reaction coordinates. Recent methodological advances in ensemble and single-molecule techniques provide complementary access to folding kinetics, conformational heterogeneity, and energy landscapes. Importantly, these methods are increasingly capable of producing reproducible, high-resolution datasets suitable for quantitative modeling.

## 3.1 Ensemble methods have revealed key principles, but average over heterogeneities

Ensemble-based experimental techniques provide access to folding kinetics, structural changes, and thermodynamic properties across a wide range of timescales and resolutions. Through revealing key features of folding mechanisms, including cooperative behavior, the presence of intermediates, and the relationship between structure and dynamics, they offer a comprehensive but still inherently averaged description of protein folding processes.

CD spectroscopy allows monitoring changes in protein secondary structure during folding and unfolding processes[9] via measuring the differential absorption of left- and right-circularly polarized light by the peptide backbone typically in the far-UV region, where characteristic spectral signatures report on the presence of α-helices, β-sheets, and IDRs. This approach enables the determination of global folding transitions,[134] thermodynamic stability,[135] and, in favorable cases, the identification of intermediate states through deviations from two-state behavior.[136] CD measurements have contributed to establishing general principles of protein folding, particularly regarding the role of secondary structure formation and the cooperative nature of the folding process.[137]

FTIR spectroscopy provides a complementary ensemble method for monitoring changes in protein secondary structure during folding and unfolding processes.[10] The approach is based on the absorption measurements of infrared radiation by the peptide backbone, most notably in the amide I region (∼1600–1700 cm⁻¹), where distinct vibrational modes report on the presence of distinct secondary structures.[138,139] By characterizing global folding transitions, thermodynamic stability, and the detection of intermediate states, particularly those involving changes in hydrogen bonding patterns and secondary structure content, FTIR provided insights into a detailed understanding of the structural basis of both soluble and membrane protein folding, especially regarding the formation and rearrangement of secondary structure elements along the folding pathway.[140]

[134] N. J. Greenfield, “Analysis of the Kinetics of Folding of Proteins and Peptides Using Circular Dichroism,” *Nature Protocols* 1 (2006): 2891–99, https://doi.org/10.1038/nprot.2006.244.
[135] N. J. Greenfield, “Using Circular Dichroism Collected as a Function of Temperature to Determine the Thermodynamics of Protein Unfolding and Binding Interactions,” *Nature Protocols* 1 (2006): 2527–35, https://doi.org/10.1038/nprot.2006.204.
[136] J. Seelig and H. J. Schönfeld, “Thermal Protein Unfolding by Differential Scanning Calorimetry and Circular Dichroism Spectroscopy Two-State Model versus Sequential Unfolding,” *Quarterly Reviews of Biophysics* 49 (2016): 9, https://doi.org/10.1017/S0033583516000044.
[137] S. M. Kelly et al., “How to Study Proteins by Circular Dichroism,” *Biochimica et Biophysica Acta (BBA) - Proteins and Proteomics* 1751 (2005): 119–39, https://doi.org/10.1016/j.bbapap.2005.06.005.

Stopped-flow spectroscopy enabled rapid mixing experiments with millisecond resolution, allowing determination of folding and unfolding rate constants and identification of two-state versus multi-state behavior.[141] Here, rapid mixing of protein solutions with denaturants or buffer triggers folding or unfolding reactions, and the resulting changes in spectroscopic signals (such as fluorescence or circular dichroism) are monitored in real time. These experiments demonstrated that many small proteins fold cooperatively, while others populate detectable intermediates. For example, stopped-flow studies of apomyoglobin revealed the presence of partially folded states,[142] providing key evidence for the existence of structured states prior to the acquisition of the native fold. The evidence for a two-state transition in folding of chymotrypsin inhibitor has also been illustrated.[143] From these measurements, kinetic parameters such as rate constants, amplitudes, and intermediate lifetimes can be extracted, enabling the reconstruction of folding energy landscapes and the discrimination between alternative mechanistic models.

NMR spectroscopy provides residue-level structural and dynamical information.[144] This technique is based on the application of a strong magnetic field, which causes nuclear spins to precess at characteristic radio frequencies that are then detected to yield information on the local chemical environment of atoms. Relaxation dispersion experiments detect low-population excited states on microsecond-to-millisecond timescales, revealing transient intermediates that are invisible to static structure determination. NMR spectroscopy has been used to study the refolding of apomyoglobin following rapid mixing experiments,[142] the low-populated folding intermediates of Fyn SH3,[145] and the detection of low-populated excited states involved in enzyme catalysis and ligand binding in systems such as dihydrofolate reductase.[146] Real-time NMR can monitor folding under controlled perturbations, offering insights into folding pathway-dependent structural changes.[147]

[138] D. M. Byler and H. Susi, “Examination of the Secondary Structure of Proteins by Deconvolved FTIR Spectra,” *Biopolymers* 25 (1986): 469–87, https://doi.org/10.1002/bip.360250307.
[139] D. Reinstädler et al., “Refolding of Thermally and Urea-Denatured Ribonuclease A Monitored by Time-Resolved FTIR Spectroscopy,” *Biochemistry* 35 (1996): 15822–30, https://doi.org/10.1021/bi961810j.
[140] M. Mimeault and D. Bonenfant, “FTIR Spectroscopic Analyses of the Temperature and pH Influences on Stratum Corneum Lipid Phase Behaviors and Interactions,” *Talanta* 56 (2002): 395–405, https://doi.org/10.1016/S0039-9140(01)00565-3.
[141] W. A. Eaton et al., “Fast Kinetics and Mechanisms in Protein Folding,” *Annual Review of Biophysics and Biomolecular Structure* 29 (2000): 327–59, https://doi.org/10.1146/annurev.biophys.29.1.327.
[142] D. Eliezer et al., “Structural and Dynamic Characterization of Partially Folded States of Apomyoglobin and Implications for Protein Folding,” *Nature Structural Biology* 5 (1998): 148–55, https://doi.org/10.1038/nsb0298-148.
[143] S. E. Jackson and A. R. Fersht, “Folding of Chymotrypsin Inhibitor 2. 1. Evidence for a Two-State Transition,” *Biochemistry* 30 (1991): 10428–35, https://doi.org/10.1021/bi00107a010.
[144] M. Osawa et al., “Functional Dynamics of Proteins Revealed by Solution NMR,” *Current Opinion in Structural Biology* 22 (2012): 660–69, https://doi.org/10.1016/j.sbi.2012.08.007.
[145] D. M. Korzhnev et al., “Low-Populated Folding Intermediates of Fyn SH3 Characterized by Relaxation Dispersion NMR,” *Nature* 430 (2004): 586–90, https://doi.org/10.1038/nature02655.

HDX-MS maps the temporal acquisition of backbone protection during folding.[41] In this approach, the protein of interest is exposed to deuterated solvent ($D_2O$) and the rate at which backbone amide hydrogens exchange with deuterium is monitored by mass spectrometry, providing residue-level or peptide-level information on solvent accessibility and structural stability. By measuring such exchange rates, HDX identifies regions that become structured early or late in the folding process. The resulting data can be used to determine the orientation of proteins within macromolecular assemblies, to assess the conformational consequences of interactions, and to characterize these interactions from both thermodynamic and kinetic perspectives. For example, H/D mapping has been applied to localize the binding site of a small-molecule inhibitor of the Eg5 kinesin motor domain,[148] to characterize IgG1 conformation and conformational dynamics,[149] to study the tyrosine Phosphorylation in the LRP1 Cytoplasmic Domain.[150] These measurements have supported the concept of cooperative structural units[151] (foldons) and provided evidence for hierarchical assembly in certain systems.[152]

Furthermore, state-of-the-art X-ray sources like X-ray FELS and/or 4$^{th}$ generation synchrotrons still provide average ensemble information.[153,154] Small-angle X-ray scattering (SAXS) determines the size and overall shape of an ensemble of protein molecules in solution. In this technique, X-rays scattered at small angles by proteins in solution are recorded, and the resulting intensity profile is used to infer their size, shape, and global structural features. It was also shown that it can contribute on the study of intermediate as well as conformational ensembles in intrinsically disordered proteins (IDPs).[155]Time-resolved

[146] D. D. Boehr et al., “The Dynamic Energy Landscape of Dihydrofolate Reductase Catalysis,” *Science* 313 (2006): 1638–42, https://doi.org/10.1126/science.1130258.
[147] G. Pintér et al., “Real-Time Nuclear Magnetic Resonance Spectroscopy in the Study of Biomolecular Kinetics and Dynamics,” *Magnetic Resonance* 2 (2021): 291–320, https://doi.org/10.5194/mr-2-291-2021.
[148] S. Brier et al., “Identification of the Protein Binding Region of S -Trityl- L -Cysteine, a New Potent Inhibitor of the Mitotic Kinesin Eg5,” *Biochemistry* 43 (2004): 13072–82, https://doi.org/10.1021/bi049264e.
[149] D. Houde et al., “Characterization of IgG1 Conformation and Conformational Dynamics by Hydrogen/Deuterium Exchange Mass Spectrometry,” *Analytical Chemistry* 81 (2009): 2644–51, https://doi.org/10.1021/ac802575y.
[150] G. N. Betts et al., “Structural and Functional Consequences of Tyrosine Phosphorylation in the LRP1 Cytoplasmic Domain,” *Journal of Biological Chemistry* 283 (2008): 15656–64, https://doi.org/10.1074/jbc.M709514200.
[151] Y. Bai et al., “Protein Folding Intermediates: Native-State Hydrogen Exchange,” *Science* 269 (1995): 192–97, https://doi.org/10.1126/science.7618079.
[152] S. W. Englander et al., “Protein Folding—How and Why: By Hydrogen Exchange, Fragment Separation, and Mass Spectrometry,” *Annual Review of Biophysics* 45 (2016): 135–52, https://doi.org/10.1146/annurev-biophys-062215-011121.
[153] K. Ayyer et al., “Perspectives for Imaging Single Protein Molecules with the Present Design of the European XFEL,” *Structural Dynamics* 2 (2015): 041702, https://doi.org/10.1063/1.4919301.
[154] J. R. Helliwell and E. P. Mitchell, “Synchrotron Radiation Macromolecular Crystallography: Science and Spin-Offs,” *IUCrJ* 2 (2015): 283–91, https://doi.org/10.1107/S205225251402795X.
[155] S. Naudi-Fabra et al., “Quantitative Description of Intrinsically Disordered Proteins Using Single-Molecule FRET, NMR, and SAXS,” *Journal of the American Chemical Society* 143 (2021): 20109–21, https://doi.org/10.1021/jacs.1c06264.

X-ray solution scattering (TR-XSS)[156,157,158] has been developed to observe changes in this ensemble average with time following an external perturbation, such as a light pulse, a heating impulse, or the introduction of chemical reactants. This method addresses the dynamics of the proteins by directly measuring the number of global conformational intermediates along a reaction trajectory, the time-scale with which they rise and decay, and hypotheses regarding the structural trajectory may be supported or ruled out. Time-resolved X-ray solution scattering using synchrotron radiation has also been used to study the refolding of cytochrome c following the photodissociation of a ligand from its heme group,[156] the folding of a pH sensitive oligonucleotide in response to a pH jump,[159] the unfolding of apomyoglobin following a heat impulse,[160] the unfolding of an α-helix of a light-oxygen-voltage protein photosensory domain.[161] SAXS studies of staphylococcal nuclease when exposed to variations in temperature, pressure, and other denaturing agents also provide structural information on unfolded states.[159] From both approaches, parameters such as the radius of gyration of the ensemble average allow the native, molten globule, and unfolded states to be characterized. The correlation of TR-XSS and time-resolved fluorescence energy transfer (TR-FRET) data has been proposed as a tool to characterize barrier-limited chain collapses on the microsecond range.[17]

Neutron scattering provides structural information on proteins in solution under near-physiological conditions, enabling the characterization of conformational ensembles with minimal radiation damage and allowing to follow both folded and partially folded conformations that may be inaccessible to X-rays. Indeed, by directing a beam of neutrons at a protein solution and measuring how the neutrons are scattered by atomic nuclei, neutron scattering provides a unique sensitivity to light atoms such as hydrogen and enables isotopic contrast variation, allowing selective visualization of specific components within complex biological systems. Unlike crystallographic techniques, which select for well-folded and stable states, solution neutron methods such as small-angle neutron scattering (SANS)[162] capture intermediate, heterogeneous, and transient conformational ensembles by resolving their distinct scattering signatures, making them particularly well-suited for studying protein folding mechanisms. Moreover, neutron scattering enables isotopic contrast variation through selective deuteration, whereby adjusting the $H_2O/D_2O$ ratio can “match out” specific components, e.g., crowding agents, effectively isolating the protein signal and revealing how environmental factors influence folding pathways. This makes neutron scattering suited for studying how proteins respond to denaturants, pressure, pH, and other perturbations.[163] Overall, the ability of neutrons to distinguish hydrogen from deuterium, to penetrate complex environments within a sample, and to capture subtle structural rearrangements without forcing proteins into rigid crystalline forms make neutron scattering one of the most insightful tools for unraveling the molecular principles that govern protein folding. For example, neutron scattering has been used to investigate surfactant-induced protein unfolding,[164] protein folding in membranes,[165] and protein adaptation to extreme temperature.[166] Neutron reflectometry further complements these insights by resolving how folding and unfolding events alter the structure of protein interfaces, such as membranes, surfaces, or solvent boundaries, providing depth-resolved information that cannot be accessed through bulk scattering alone.[167]

[156] M. Cammarata et al., “Tracking the Structural Dynamics of Proteins in Solution Using Time-Resolved Wide-Angle X-Ray Scattering,” *Nature Methods* 5 (2008): 881–86, https://doi.org/10.1038/nmeth.1255.
[157] M. Andersson et al., “Structural Dynamics of Light-Driven Proton Pumps,” *Structure* 17 (2009): 1265–75, https://doi.org/10.1016/j.str.2009.07.007.
[158] Y. Lee et al., “A Comparative Review of Time-Resolved x-Ray and Electron Scattering to Probe Structural Dynamics,” *Structural Dynamics* 11 (2024): 031301, https://doi.org/10.1063/4.0000249.
[159] A. M. Chan et al., “Early Folding Dynamics of I-Motif DNA Revealed by pH-Jump Time-Resolved X-Ray Solution Scattering,” *Journal of the American Chemical Society* 146 (2024): 33743–52, https://doi.org/10.1021/jacs.4c11768.
[160] L. Henry et al., “Real-Time Tracking of Protein Unfolding with Time-Resolved x-Ray Solution Scattering,” *Structural Dynamics* 7 (2020): 054702, https://doi.org/10.1063/4.0000013.
[161] P. E. Konold et al., “Microsecond Time-Resolved X-Ray Scattering by Utilizing MHz Repetition Rate at Second-Generation XFELs,” *Nature Methods* 21 (2024): 1608–11, https://doi.org/10.1038/s41592-024-02344-0.
[162] V. K. Aswal et al., “Small-Angle Neutron Scattering Study of Protein Unfolding and Refolding,” *Physical Review E* 80 (2009): 011924, https://doi.org/10.1103/PhysRevE.80.011924.

Cryo-EM resolves the 3D structure of a macromolecule by reconstructing images of vitrified molecules, enabling characterization of conformational states within heterogeneous ensembles.[168] The method reconstructs the structure of a molecule by computationally averaging large populations of individual nearly identical proteins, typically tens of thousands of images. This averaging process is essential to improve the signal-to-noise ratio and enables the reconstruction of high-resolution 3D structures of proteins and protein complexes. Importantly, modern image-processing approaches allow the separation of molecular subpopulations, visualizing structural heterogeneity within a sample from both *in vitro*[169] and, more complicated and less developed, in situ samples.[170] However, the achievable resolution of each conformational state strongly depends on its relative abundance and stability, as less populated states provide lower statistical power for reconstruction. As a result, cryo-EM often yields ensembles of structures that reflect the conformational landscape of a macromolecular system rather than a single static conformation, but still represents an equilibrium structural state of the analyzed protein. In recent years, time-resolved cryo-EM approaches have begun to emerge,[171] aiming to capture transient intermediates along functional reaction pathways by rapidly initiating reactions and vitrifying or melting samples at defined time points. Although still technically challenging, these strategies promise to extend cryo-EM from a primarily equilibrium technique toward direct visualization of molecular dynamics. However, this technique cannot directly resolve how the detected states are connected to each other; in other words, folding pathways can be inferred but not observed. At the same time, the rapid expansion of cryo-EM structural data has revealed a remarkable level of structural polymorphism in biological assemblies. For example, cryo-EM has played a central role in demonstrating that a single protein sequence can give rise to structurally diverse aggregates depending on environmental conditions or biological context with potentially different biological or pathological properties, as it happens for amyloid fibrils.[172] Together, cryo-EM is a powerful method capable of describing not only high-resolution structures but also the broader structural heterogeneity and conformational landscapes that characterize many biological macromolecules.

---

[163] A. Paciaroni et al., “Effect of the Environment on the Protein Dynamical Transition: A Neutron Scattering Study,” *Biophysical Journal* 83 (2002): 1157–64, https://doi.org/10.1016/S0006-3495(02)75239-9.
[164] S. Chodankar et al., “Surfactant-Induced Protein Unfolding as Studied by Small-Angle Neutron Scattering and Dynamic Light Scattering,” *Journal of Physics: Condensed Matter* 19 (August 2007): 326102, https://doi.org/10.1088/0953-8984/19/32/326102.
[165] X. Han et al., “Protein Folding in Membranes: Insights from Neutron Diffraction Studies of a Membrane β-Sheet Oligomer,” *Biophysical Journal* 94 (2008): 492–505, https://doi.org/10.1529/biophysj.107.113183.
[166] M. Tehei et al., “Neutron Scattering Reveals the Dynamic Basis of Protein Adaptation to Extreme Temperature,” *Journal of Biological Chemistry* 280 (2005): 40974–79, https://doi.org/10.1074/jbc.M508417200.
[167] G. Fragneto-Cusani, “Neutron Reflectivity at the Solid/Liquid Interface: Examples of Applications in Biophysics,” *Journal of Physics: Condensed Matter* 13 (2001): 4973–89, https://doi.org/10.1088/0953-8984/13/21/322.
[168] X. Bai et al., “How Cryo-EM Is Revolutionizing Structural Biology,” *Trends in Biochemical Sciences* 40 (2015): 49–57, https://doi.org/10.1016/j.tibs.2014.10.005.
[169] D. Kimanius and J. Schwab, “Confronting Heterogeneity in Cryogenic Electron Microscopy Data: Innovative Strategies and Future Perspectives with Data-Driven Methods,” *Current Opinion in Structural Biology* 86 (2024): 102815, https://doi.org/10.1016/j.sbi.2024.102815.
[170] A. Bartesaghi, “Strategies for Studying Discrete Heterogeneity in Situ Using Cryo-Electron Tomography,” *Current Opinion in Structural Biology* 95 (2025): 103186, https://doi.org/10.1016/j.sbi.2025.103186.
[171] U. J. Lorenz, “Advances in Microsecond Time-Resolved Cryo-EM,” *Structural Dynamics* 12 (2025): 23–23, https://doi.org/10.1063/4.0000815.

Despite their strengths, ensemble approaches still report ensemble-average behavior. When multiple folding pathways coexist, ensemble signals may obscure heterogeneity and reduce complex kinetic networks to effective rate constants.[37] While methods such as cryo-EM can partially recover structural heterogeneity, they still rely on averaging within each class and do not directly resolve the dynamical connectivity between states. As a result, complex kinetic networks are often reduced to effective, averaged rate constants, and low-population intermediates or rare transition events, critical for understanding folding mechanisms, remain undetected. This limitation becomes particularly pronounced for proteins with rugged energy landscapes or competing pathways, where ensemble measurements may mask stochastic variability and hinder a complete mechanistic description of the folding process.

## 3.2 Single-molecule fluorescence spectroscopy exposes stochastic transitions between conformations

Single-molecule fluorescence methods circumvent ensemble averaging by directly resolving the structural dynamics of individual molecules.[173,174] Super-resolution fluorescence microscopy provides a powerful approach for overcoming the diffraction limit of conventional light microscopy and resolving structural heterogeneity at the nanometer scale. Techniques like slimfield or narrowfield microscopy operate based on the size reduction of the illumination area enabling tracking of individual fluorescent proteins at a millisecond time scale. This in turn reduces the motion blur of diffusing fluorescently labelled particles. This is particularly relevant when visualizing proteins directly in living cells due to the number of different cellular compartments where target proteins can be localized and, despite being a highly crowded environment, high protein diffusion rates in the cytoplasm.[175,176] Other strategies include photoactivated localization microscopy (PALM) and stochastic optical reconstruction microscopy (STORM) or targeted depletion of fluorescence emission (e.g., STED), enabling the precise determination of molecular positions beyond the ~200 nm limit of diffraction with a spatial resolution that can reach 20 nm.[177] By labeling proteins or specific sites within proteins, super-resolution methods allow the visualization of conformational distributions, spatial organization, and structural rearrangements that occur during folding and assembly processes *in vitro*[178] or directly in living cells.[179] Time-resolved implementations of super-resolution techniques can capture dynamic transitions and rare conformational states, providing access to heterogeneous pathways that are not observable in ensemble measurements. These techniques have been widely applied to investigate protein organization and assembly in complex environments, including membranes in living cells,[180,181,182] and to probe the interplay between folding, misfolding, and aggregation,[183] as well as physiological oligomerization[69,184] at high spatial resolution. As such, super-resolution fluorescence microscopy complements established single-molecule approaches by linking structural dynamics to spatial and often biological context, contributing to a more comprehensive understanding of protein folding and function in physiologically relevant settings.

[172] S. H. W. Scheres et al., "Molecular pathology of neurodegenerative diseases by cryo-EM of amyloids," *Nature* 621 (2023): 701–10, https://doi.org/10.1038/s41586-023-06437-2.
[173] S. Shashkova and M. C. Leake, "Single-Molecule Fluorescence Microscopy Review: Shedding New Light on Old Problems," *Bioscience Reports* 37 (2017): 20170031, https://doi.org/10.1042/BSR20170031.
[174] S. Shashkova and M. C. Leake, "Systems Biophysics: Single-Molecule Optical Proteomics in Single Living Cells," *Current Opinion in Systems Biology* 7 (2018): 26–35, https://doi.org/10.1016/j.coisb.2017.11.006.

Another technique, single-molecule Förster resonance energy transfer (smFRET)[185,186] monitors distance fluctuations between site-specific labeled residues, revealing stochastic transitions between conformational states in real time.[187] This approach enables direct observation of folding and unfolding dynamics at the level of individual proteins. Time traces obtained from smFRET reveal discrete transitions between conformational states, allowing the determination of state lifetimes and transition probabilities. Such measurements have demonstrated that proteins previously classified as simple two-state folders may exhibit transient intermediates. In IDPs, fluorescence-based approaches instead reveal broad conformational distributions and rapid fluctuations within dynamic ensembles.[188] Together, single-molecule fluorescence provides direct access to conformational heterogeneity and stochastic transitions, revealing the diverse folding pathways and dynamic fluctuations that remain hidden in ensemble-averaged measurements.[189]

---

[175] T. Kühn et al., "Protein Diffusion in Mammalian Cell Cytoplasm," *PLOS ONE* 6 (2011): 22962, https://doi.org/10.1371/journal.pone.0022962.
[176] A. J. Wollman and M. C. Leake, "Millisecond Single-Molecule Localization Microscopy Combined with Convolution Analysis and Automated Image Segmentation to Determine Protein Concentrations in Complexly Structured, Functional Cells, One Cell at a Time," *Faraday Discussions* 184 (2015): 401–24, https://doi.org/10.1039/C5FD00077G.
[177] S. J. Sahl and W. Moerner, "Super-Resolution Fluorescence Imaging with Single Molecules," *Current Opinion in Structural Biology* 23 (2013): 778–87, https://doi.org/10.1016/j.sbi.2013.07.010.
[178] F. Zosel et al., "Labeling of Proteins for Single-Molecule Fluorescence Spectroscopy," in *Protein Folding. Methods in Molecular Biology*, vol. 2376, ed. V. Muñoz (Humana, 2022), https://doi.org/10.1007/978-1-0716-1716-8_12.
[179] S. Shashkova et al., "Copy Number Analysis of the Yeast Histone Deacetylase Complex Component Cti6 Directly in Living Cells," *Methods in Molecular Biology* (New York) 2476 (2022), https://doi.org/10.1007/978-1-0716-2221-6_14.
[180] S. Manley et al., "High-Density Mapping of Single-Molecule Trajectories with Photoactivated Localization Microscopy," *Nature Methods* 5 (2008): 155–57, https://doi.org/10.1038/nmeth.1176.
[181] M. B. Stone et al., "Super-Resolution Microscopy: Shedding Light on the Cellular Plasma Membrane," *Chemical Reviews* 117 (2017): 7457–77, https://doi.org/10.1021/acs.chemrev.6b00716.
[182] S. Shashkova et al., "Correlating Single-Molecule Characteristics of the Yeast Aquaglyceroporin Fps1 with Environmental Perturbations Directly in Living Cells," *Methods* 193 (2021): 46–53, https://doi.org/10.1016/j.ymeth.2020.05.003.
[183] D. Pinotsi et al., "Direct Observation of Heterogeneous Amyloid Fibril Growth Kinetics via Two-Color Super-Resolution Microscopy," *Nano Letters* 14 (2014): 339–45, https://doi.org/10.1021/nl4041093.
[184] J. Zalejski et al., "Unravelling the Mystery inside Cells by Using Single-Molecule Fluorescence Imaging," *Journal of Imaging* 9 (2023): 192, https://doi.org/10.3390/jimaging9090192.
[185] L. Stryer and R. P. Haugland, "Energy Transfer: A Spectroscopic Ruler," *Proceedings of the National Academy of Sciences of the United States of America* 58 (1967): 719–26, https://doi.org/10.1073/pnas.58.2.719.
[186] T. Ha et al., "Probing the Interaction between Two Single Molecules: Fluorescence Resonance Energy Transfer between a Single Donor and a Single Acceptor," *Proceedings of the National Academy of Sciences of the United States of America* 93 (1996): 6264–68, https://doi.org/10.1073/pnas.93.13.6264.

An important limitation in smFRET is the limited observation time for molecules that diffuse freely in solution. In an ordinary fluorescence correlation spectroscopy FRET (FCS-FRET) setup, the typical observation time is limited to ~1 ms as individual proteins diffuse through the diffraction limited probing volume. To observe dynamics on longer timescales, one must resort to the ensemble approach by combining data from multiple events. This has led to intense research efforts focused on confining proteins in order to extend the observation time. One possible strategy is surface tethering, but care has to be taken to minimize potential effects on the conformational freedom of the protein. Other approaches are based on optical tweezers or electrokinetic forces. However, relatively strong forces are required to overcome Brownian motion for objects as small as proteins, which often makes these approaches invasive as well. For instance, field- and/or heat induced protein denaturation has been observed already long ago, both in metallic apertures[190] and in nanopores.[191] Here, an interesting alternative is field-free trapping in containers, such as lipid vesicles. However, the capture yield is typically low, and liquid exchange becomes more difficult when the protein is surrounded by a membrane barrier. In this context, solid-state nanostructures modified with polymer brushes are an interesting alternative since they allow passage of small molecules while proteins remain securely trapped in solution phase under physiological buffer conditions.[192] Another option is convex lens-induced confinement,[193] where slightly larger solid-state containers are used to image single-molecule diffusion.

While smFRET has been a foundational tool for monitoring distance fluctuations, it provides distances rather than absolute spatial coordinates. A complementary advance for bridging this gap is MINFLUX (Minimal Photon Fluxes).[194] Unlike traditional super-resolution techniques that maximize photon collection to find a center of a molecule, MINFLUX probes the emitter with a donut-shaped laser beam to locate its dark center. This approach requires vastly fewer photons for localization, which translates directly into high temporal resolution. This capability was recently demonstrated by tracking the mechanistic steps of kinesin motor proteins with nanometer precision and sub-millisecond resolution.[124,195] For the study of protein folding, MINFLUX could therefore complement smFRET by allowing researchers to move beyond distance-based observables and, in favorable cases, track the actual vector of movement as a protein transitions from an unfolded coil to its native structure in real-time.

---

[187] E. Lerner et al., "Toward Dynamic Structural Biology: Two Decades of Single-Molecule Förster Resonance Energy Transfer," *Science* 359 (2018): eaan1133, https://doi.org/10.1126/science.aan1133.
[188] Lauren Ann Metskas and Elizabeth Rhoades, "Single-Molecule FRET of Intrinsically Disordered Proteins," *Annual Review of Physical Chemistry* 71 (2020): 391–414, https://doi.org/10.1146/annurev-physchem-012420-104917.
[189] M. Pirchi et al., "Single-molecule fluorescence spectroscopy maps the folding landscape of a large protein," *Nature Communications* 2 (2011): 493, https://doi.org/10.1038/ncomms1504.
[190] Y. Pang and R. Gordon, "Optical Trapping of a Single Protein," *Nano Letters* 12 (2012): 402–6, https://doi.org/10.1021/nl203719v.
[191] D. S. Talaga and J. Li, "Single-Molecule Protein Unfolding in Solid State Nanopores," *Journal of the American Chemical Society* 131 (2009): 9287–97, https://doi.org/10.1021/ja901088b.
[192] J. Svirelis et al., "Stable Trapping of Multiple Proteins at Physiological Conditions Using Nanoscale Chambers with Macromolecular Gates," *Nature Communications* 14 (2023): 5131, https://doi.org/10.1038/s41467-023-40889-4.
[193] S. Leslie et al., "Single-Molecule Imaging of the Biophysics of Molecular Interactions with Precision and Control, in Cell-like Conditions, and without Tethers," *Current Opinion in Biomedical Engineering* 12 (2019): 75–82, https://doi.org/10.1016/j.cobme.2019.10.004.

The threshold of structural biology has been further crossed by the development of RESI (Resolution Enhancement by Sequential Imaging[196]), which acts as an “Ångström ruler” for static structures. While MINFLUX is the preferred tool for dynamics and fast conformational changes, RESI offers the ultimate spatial resolution for measuring intramolecular bond distances in fixed states. By utilizing sequential barcoding of DNA imager strands, RESI can distinguish molecules that are chemically identical and spatially overlapping, providing resolution limited only by the label size. This has been applied to resolve the subunits of the Nuclear Pore Complex (Nup96) and CD20 dimers with <1 nm resolution. Together, these technologies provide a dual approach to protein folding: MINFLUX for folding dynamics and RESI for the high precision of fixed structural states.

As optical resolution reaches the Ångström scale, the critical bottleneck for structural biology shifts from the physics of the microscope to the chemistry of labeling, a challenge known as linkage error. Linkage error refers to the physical displacement between the target molecule and its fluorescent label. Immunolabeling with standard primary–secondary antibody sandwiches, a traditional approach in structural biology, can add a ~20 nm displacement, which effectively masks the 1 nm precision possible by MINFLUX or RESI. To translate optical resolution into biological insight, it is necessary to move toward smaller labels such as nanobodies (~3 nm) or direct labeling strategies. Endogenous fluorescent protein fusions have been widely used for protein studies in cells. More recent advances including incorporation of non-canonical amino acids directly to the protein of interest backbone in a selective biorthogonal manner[197] significantly improve experimental precision for both *in vitro* and *in vivo* protein studies, enabling sub-nanometer linkage and allowing the microscope’s precision to match the biological scale of the protein structure.

Advances in photon-detection technologies, fluorescence instrumentation, and computational analysis have improved the temporal resolution, signal-to-noise ratio, and throughput of single-molecule fluorescence measurements. The latter has been recently enhanced by multiplexed single-molecule approaches by enabling measurements across large molecular libraries. By linking single-molecule observations with sequencing-based identification, approaches such as MUSCLE,[198] SPARXS,[199] and Spin-Seq[200] allow dynamic and structural measurements to be systematically mapped across sequence space, bridging mechanistic single-molecule biophysics with high-throughput molecular screening.[201] However, applications to date have largely focused on nucleic-acid sequence libraries, including studies of protein-nucleic acid interactions.[198,200] Analogous measurements using DNA-barcoded peptide or protein libraries are conceptually possible, but remain unexplored.

---

194 Francisco Balzarotti et al., “Nanometer Resolution Imaging and Tracking of Fluorescent Molecules with Minimal Photon Fluxes,” *Science* 355 (February 2017): 606–12, https://doi.org/10.1126/science.aak9913.
195 Jan O. Wolff et al., “MINFLUX Dissects the Unimpeded Walking of Kinesin-1,” *Science* 379 (March 2023): 1004–10, https://doi.org/10.1126/science.ade2650.
196 Susanne C. M. Reinhardt et al., “Ångström-Resolution Fluorescence Microscopy,” *Nature* 617 (May 2023): 711–16, https://doi.org/10.1038/s41586-023-05925-9.
197 L. M. Silbermann et al., “One-Pot Dual Protein Labeling for Simultaneous Mechanical and Fluorescent Readouts in Optical Tweezers,” *Protein Science* 34 (2025): e70098, https://doi.org/10.1002/pro.70098.

Despite field advances, several limitations remain. For freely diffusing molecules, observation times are limited by the short residence time within the detection volume, whereas immobilized or confined molecules are ultimately limited by fluorophore photobleaching. In addition, most single-molecule fluorescence measurements report on a limited number of selected inter-residue distances rather than providing a complete structural ensemble. Measurements in cellular environments face further challenges, including limited optical accessibility, molecular crowding, and background autofluorescence, all of which can reduce spatial precision and temporal resolution. Finally, the attachment of fluorescent labels may itself perturb protein structure or dynamics[202] and therefore requires careful validation.

## 3.3 Single-molecule force spectroscopy directly probes energy landscapes and intermediates

Single-molecule force spectroscopy, using for example optical tweezers, magnetic tweezers, atomic force microscopy (AFM), or a range of correlative approaches,[39, 203] provides a complementary strategy for interrogating folding and unfolding dynamics. By applying controlled mechanical forces to individual protein molecules, these techniques perturb the energy landscape and measure extension changes associated with conformational transitions.

Force–extension measurements and force-clamp experiments yield quantitative information about unfolding forces, refolding kinetics, intermediate states, and energy barrier heights that a polypeptide chain must overcome to reach the corresponding protein native state.[20] By varying the applied force, it is possible to map the dependence of transition rates on mechanical load and reconstruct one-dimensional projections of the underlying free-energy landscape. Transition state distances and barrier parameters can be extracted from the force dependence of unfolding and refolding rates.

---

[198] J. Aguirre Rivera et al., "Massively Parallel Analysis of Single-Molecule Dynamics on next-Generation Sequencing Chips," *Science* 385 (2024): 892–98, https://doi.org/10.1126/science.adn5371.
[199] I. Severins et al., "Single-Molecule Structural and Kinetic Studies across Sequence Space," *Science* 385 (2024): 898–904, https://doi.org/10.1126/science.adn5968.
[200] Jagadish Prasad Hazra et al., "Unraveling Single-Molecule Reactions via Multiplexed in-Situ DNA Sequencing," preprint, Biophysics, December 2, 2025, https://doi.org/10.1101/2025.11.27.690701.
[201] A. N. Kapanidis et al., "From Sequence to Function: Bridging Single-Molecule Kinetics and Molecular Diversity," *Science* 391 (2026): 458–65, https://doi.org/10.1126/science.adv4503.
[202] K. L. Schneider et al., "Comparison of Endogenously Expressed Fluorescent Protein Fusions Behaviour for Protein Quality Control and Cellular Ageing Research," *Scientific Reports* 11 (2021): 12819, https://doi.org/10.1038/s41598-021-92249-1.
[203] J. W. Shepherd et al., "Correlating Fluorescence Microscopy, Optical and Magnetic Tweezers to Study Single Chiral Biopolymers Such as DNA," *Nature Communications* 15 (2024): 2748, https://doi.org/10.1038/s41467-024-47126-6.

Single-molecule force spectroscopy has revealed discrete intermediates in proteins that appear cooperative in ensemble measurements.[204] It has also provided insights into mechanical stability in load-bearing proteins and into non-equilibrium folding behavior under controlled perturbations.[205] Because each molecule is monitored individually, stochastic variability and rare transitions are directly observable.

Optical tweezers are one of the most widely used forms of single-molecule force spectroscopy, providing precise and continuous control over the forces applied to individual protein molecules.[206] In this approach, a protein is tethered between micron-sized dielectric beads via molecular handles, with force exerted by trapping the beads in highly focused laser beams. Changes in molecular extension are monitored with nanometer precision as a function of the applied force, enabling real-time observation of folding and unfolding transitions. In both force-ramp and force-clamp modes, optical tweezers provide direct measurements of unfolding forces, refolding kinetics and the lifetimes of intermediate states, as well as the position and height of energy barriers along the folding pathway. By systematically varying the applied load, optical tweezers can reconstruct the force-dependent free energy landscape and extract transition state parameters, thus complementing the general framework described above. For instance, optical tweezers experiments have been employed to characterize the stepwise unfolding of titin domains, probe the folding pathways of small globular proteins[207] and quantify misfolding trajectories in proteins related to diseases.[208]

Magnetic tweezers provide a complementary single-molecule force spectroscopy approach in which forces are applied to individual protein molecules through magnetic fields, enabling highly stable and long-duration measurements of folding and unfolding dynamics.[209] A protein is tethered between a surface and a paramagnetic bead, and force is exerted by external magnets that generate a controlled magnetic field gradient. By tracking the bead position with nanometer resolution, changes in molecular extension associated with conformational transitions can be monitored in real time. This technique offers exceptional force stability and the ability to maintain constant force over extended timescales, which is particularly advantageous for probing slow folding kinetics and rare events. In addition, the experimental configuration allows the simultaneous manipulation of many molecules, enabling high-throughput measurements and improved statistical sampling. This method allows precise determination of unfolding and refolding rates, equilibrium folding probabilities, and the force dependence of transition kinetics. Magnetic tweezers have been used, for example, to investigate the folding dynamics of small protein domains,[210] to quantify the effects of chaperones on folding under force,[211] and to explore the role of mechanical constraints in modulating protein stability.[212]

---

[204] D. R. Jacobson, "Single Molecule Force Spectroscopy to Probe Intermediates and Energetics of Membrane Protein Folding," *Chemical Reviews*, ahead of print, 2026, https://doi.org/10.1021/acs.chemrev.5c00612.
[205] H. Sun et al., "Protein Folding Mechanism Revealed by Single-Molecule Force Spectroscopy Experiments," *Biophysics Reports* 7 (2021): 399, https://doi.org/10.52601/bpr.2021.210024.
[206] C. J. Bustamante et al., "Optical tweezers in single-molecule biophysics," *Nature Reviews Methods Primers* 1 (2021): 25, https://doi.org/10.1038/s43586-021-00021-6.
[207] C. Cecconi et al., "Direct Observation of the Three-State Folding of a Single Protein Molecule," *Science* 309 (2005): 2057–60, https://doi.org/10.1126/science.1116702.
[208] P. O. Heidarsson et al., "Direct Single-Molecule Observation of Calcium-Dependent Misfolding in Human Neuronal Calcium Sensor-1," *Proceedings of the National Academy of Sciences of the United States of America* 111 (2014): 13069–74, https://doi.org/10.1073/pnas.1401065111.
[209] I. D. Vilfan et al., "Magnetic Tweezers for Single-Molecule Experiments," in *Handbook of Single-Molecule Biophysics*, ed. P. Hinterdorfer and A. Oijen (Springer US, 2009), https://doi.org/10.1007/978-0-387-76497-9_13.

AFM is another single-molecule force spectroscopy approach in which mechanical force is applied to proteins by physically pulling on them with a nanometer-scale cantilever, enabling direct measurement of unfolding and refolding processes.[213] A protein is typically attached between a substrate and the tip of a flexible cantilever, and force is exerted by retracting the cantilever. Changes in molecular extension are recorded as force–extension curves, in which discrete unfolding events appear as characteristic steps that correspond to the sequential unfolding of structural domains. AFM allows the determination of unfolding forces, contour length changes, and the mechanical stability of individual domains, as well as the identification of intermediate states along the folding pathway. AFM-based force spectroscopy has been widely used to characterize the stepwise unfolding of modular proteins such as titin,[214] to investigate the mechanical properties of membrane and cytoskeletal proteins,[215] and to probe the effects of mutations or ligand binding on protein stability.[216]

Recent improvements in instrumentation, including enhanced force stability, multiplexing strategies,[217] and automated feedback control,[218] are increasing experimental throughput and reproducibility. These developments make it possible to acquire statistically robust datasets of folding and unfolding trajectories under systematically varied conditions.[219]

---

[210] H. K. Choi et al., "High-Resolution Single-Molecule Magnetic Tweezers," *Annual Review of Biochemistry* 91 (2022): 33–59, https://doi.org/10.1146/annurev-biochem-032620-104637.
[211] S. Haldar et al., "Trigger Factor Chaperone Acts as a Mechanical Foldase," *Nature Communications* 8 (2017): 668, https://doi.org/10.1038/s41467-017-00771-6.
[212] R. Tapia-Rojo et al., "Single-Molecule Magnetic Tweezers to Probe the Equilibrium Dynamics of Individual Proteins at Physiologically Relevant Forces and Timescales," *Nature Protocols* 19 (2024): 1779–806, https://doi.org/10.1038/s41596-024-00965-5.
[213] J. Zlatanova et al., "Single Molecule Force Spectroscopy in Biology Using the Atomic Force Microscope," *Progress in Biophysics and Molecular Biology* 74 (2000): 37–61, https://doi.org/10.1016/S0079-6107(00)00014-6.
[214] M. Rief et al., "Reversible Unfolding of Individual Titin Immunoglobulin Domains by AFM," *Science* 276 (1997): 1109–12, https://doi.org/10.1126/science.276.5315.1109.
[215] D. J. Müller and A. Engel, "Atomic Force Microscopy and Spectroscopy of Native Membrane Proteins," *Nature Protocols* 2 (2007): 2191–97, https://doi.org/10.1038/nprot.2007.309.
[216] R. Ros et al., "Single Molecule Force Spectroscopy on Ligand–DNA Complexes: From Molecular Binding Mechanisms to Biosensor Applications," *Journal of Biotechnology* 112 (2004): 5–12, https://doi.org/10.1016/j.jbiotec.2004.04.029.
[217] Yannic Kerkhoff et al., "Microfluidics-Based Force Spectroscopy Enables High-Throughput Force Experiments with Sub-Nanometer Resolution and Sub-Piconewton Sensitivity," *Small* 19 (April 2023): 2206713, https://doi.org/10.1002/smll.202206713.
[218] Martin Selin et al., "SmartTrap: Automated Precision Experiments with Optical Tweezers," version 1, preprint, arXiv, 2025, https://doi.org/10.48550/ARXIV.2505.05290.
[219] S. A. Kim et al., "Membrane Protein Folding and Biogenesis: Insights from Single-Molecule Force Spectroscopy," *Chemical Reviews*, ahead of print, 2026, https://doi.org/10.1021/acs.chemrev.5c01097.

## 3.4 Systematic acquisition of quantitative folding trajectories is now becoming feasible

The technical landscape is shifting from isolated case studies toward systematic data generation. Table 1 summarizes the principal experimental approaches, their accessible timescales and spatial resolutions, and their respective strengths and limitations. Together, these techniques provide complementary constraints on folding pathways, conformational distributions, and energy landscapes. Emerging approaches such as time-resolved X-ray free-electron laser (XFEL) experiments[220] enable ultrafast “molecular movies” of biomolecular dynamics and may, in the future, provide complementary insights into folding-related conformational changes.

Advances in microfluidics, instrument control, and automation allow repeated and standardized measurements across multiple molecules and conditions. Multiplexed force spectroscopy setups can monitor several tethers in parallel,[221] while automated analysis pipelines reduce subjective intervention in state identification and kinetic fitting.[222] Similarly, improvements in fluorescence detection and labeling strategies support larger-scale acquisition of single-molecule trajectories. Standardized data formats and reproducible experimental protocols facilitate comparison across laboratories and systems. The convergence of ensemble assays, single-molecule fluorescence, and force-based methods creates an opportunity to assemble quantitative datasets of folding trajectories under defined perturbations.

A predictive framework linking sequence to folding mechanisms will require datasets that capture not only native structures, but also intermediate populations, transition kinetics, and response to environmental perturbations. The experimental tools to generate such trajectory-level information are now sufficiently mature to support this objective.

---

[220] J. Moon et al., “Time-Resolved Serial Femtosecond Crystallography for Investigating Structural Dynamics of Chemical Systems,” *Chemical Communications* 60 (2024): 9472–82, https://doi.org/10.1039/D4CC03185G.
[221] A. Löf et al., “Multiplexed Protein Force Spectroscopy Reveals Equilibrium Protein Folding Dynamics and the Low-Force Response of von Willebrand Factor,” *Proceedings of the National Academy of Sciences* 116 (2019): 18798-18807, https://doi.org/10.1073/pnas.1901794116.
[222] M. Rico-Pasto et al., “Force-Dependent Folding Kinetics of Single Molecules with Multiple Intermediates and Pathways,” *The Journal of Physical Chemistry Letters* 13 (2022): 1025–32, https://doi.org/10.1021/acs.jpclett.1c03521.

**Table 1 | Experimental techniques for studying protein folding and misfolding.**

| Technique | Principle | Spatial Resolution | Temporal Resolution | What It Directly Measures | Strengths | Limitations for Folding Mechanism |
|---|---|---|---|---|---|---|
| **Circular dichroism (CD),[9] Fourier transform infrared spectroscopy (FTIR),[10] and fluorescence spectroscopy.[223]** | Detect changes in secondary structure or probe fluorescence during folding transitions. | Global or probe-dependent. | μs–s. | Secondary structure formation; global folding transitions. | Rapid kinetic measurements; accessible. | Usually no residue-level detail; ensemble averaged. |
| **Stopped-flow kinetics.[11]** | Rapidly mixes reactants and monitors spectroscopic changes during folding. | Global or probe-dependent. | ms–s. | Folding and unfolding rate constants. | Quantitative kinetics; well-established. | Lacks structural resolution; ensemble averaging. |
| **Nuclear magnetic resonance (NMR) spectroscopy.[57]** | Probes atomic environments through magnetic resonance of nuclei in solution. | Atomic to residue level. | μs–s (indirectly via relaxation). | Conformational ensembles; exchange processes; residue-level dynamics. | Probes dynamics in solution; site-specific information. | Limited to smaller proteins; ensemble averaged. |
| **Hydrogen–deuterium exchange mass spectrometry (HDX-MS).[41]** | Measures exchange of backbone amide hydrogens to infer local stability and protection. | Peptide to residue level. | ms–h. | Local stability and protection patterns. | Identifies cooperative units; applicable to large systems. | Indirect structural inference; limited temporal precision. |
| **X-ray crystallography.[15]** | Determines atomic structure from diffraction of X-rays by ordered protein crystals. | Atomic (~1–3 Å). | Static. | High-resolution 3D structure of stable states. | Atomic detail; foundation of structural databases. | Captures endpoints; limited access to transient intermediates; requires crystallization. |
| **Time-resolved X-ray solution scattering.[224]** | Provides information on the size and shape of a protein ensemble. | Sub-nanometer one-dimensional structural information. | ps–s. | Time-dependent perturbations to the average global conformation of a protein ensemble. | Easily resolves the number of global conformational states and their rate of growth and decay. | Does not provide direct resolution structural information, but rather data against which to test structural hypotheses. |
| **Small Angle Neutron Scattering.[225]** | Non-destructive low resolution structure in physiological conditions. | Molecular level. | Both static and dynamic. | Structure from 0.5 to 300nm; dynamics from ms to hr. | Isotopic substitution allows contrast matching for example of crowding agents the induce folding. Powerful application structure of membrane[226] proteins in nanodiscs. | Lower resolution compared to synchrotron methods. |
| **Cryo-electron microscopy (Cryo-EM).[227]** | Reconstructs structures from averaged images of rapidly frozen particles. | Near-atomic (2–4 Å typical). | Static snapshots. | Structures of large complexes and assemblies. | Access to large systems; multiple conformations possible. | Limited temporal information; ensemble averaging. |

[223] R. Nolan et al., “Detecting Protein Aggregation and Interaction in Live Cells: A Guide to Number and Brightness,” *Methods* 140 (2018): 172–77, https://doi.org/10.1016/j.ymeth.2017.12.001.
[224] Hyun Sun Cho et al., “Protein Structural Dynamics in Solution Unveiled via 100-Ps Time-Resolved x-Ray Scattering,” *Proceedings of the National Academy of Sciences* 107 (April 2010): 7281–86, https://doi.org/10.1073/pnas.1002951107.
[225] E. E. Lattman et al., “SANS,” in *Biological Small Angle Scattering: Theory and Practice* (Oxford University Press, 2018), https://doi.org/10.1093/oso/9780199670871.003.0011.
[226] W. Javed et al., “Structural Insights into the Catalytic Cycle of a Bacterial Multidrug ABC Efflux Pump,” *Journal of Molecular Biology* 434 (2022): 167541, https://doi.org/10.1016/j.jmb.2022.167541.
[227] H. R. Saibil, “Cryo-EM in molecular and cellular biology,” *Molecular Cell* 82 (2022): 274–84, https://doi.org/10.1016/j.molcel.2021.12.016.

| | | | | | | |
|---|---|---|---|---|---|---|
| **IR spectroscopy of isolated peptides.**[228] | Determines structures of gas-phase peptides by combing high-resolution IR spectroscopy with quantum chemical calculations. | Atomic level. | Static. | High-resolution IR spectra of jet-cooled peptides. | Structures of individual conformers, probing non-covalent inter- and intra-molecule interactions. | No temporal information, small peptides. |
| **Single-molecule FRET (smFRET).**[186] | Measures distance changes between fluorescently labeled sites on individual molecules. | Nanometer scale. | μs–s, depending whether a confinement method used. | Distance changes between labeled sites; state transitions. | Resolves heterogeneity; real-time transitions. | Requires labeling; limited structural resolution; few observables per molecule. |
| **Single-molecule force spectroscopy (optical tweezers, AFM).**[229] | Applies controlled mechanical force to individual molecules to monitor folding and unfolding trajectories. | Nanometer projection along force axis. | μs–minutes. | Folding and unfolding trajectories; barrier heights; energy landscape projections. | Direct measurement of intermediates and transition states; access to stochastic trajectories. | One-dimensional projection; mechanical perturbation alters landscape. |
| **MINFLUX.**[194] | Probes emitter position with a local excitation minimum (donut beam). | 1–3 nm. | Sub-ms. | Real-time trajectories of protein domains and fast conformational transitions. | Exceptional temporal resolution for dynamics; requires significantly fewer photons for localization than traditional super-resolution. | Resolution is highly sensitive to “linkage error” from label size; restricted by the chemistry of fluorescent labeling. |
| **RESI.**[196] | Sequential barcoding of DNA imager strands to separate adjacent targets in time. | Ångström scale (<1 nm). | Static snapshots. | Intramolecular distances and subunits in fixed protein complexes. | Acts as an "Ångström ruler"; can distinguish spatially overlapping molecules that are chemically identical. | Lacks temporal information; cannot observe active folding pathways, only fixed structural states. |
| **Super-resolution fluorescence microscopy.**[173] | Localizes fluorescently labeled molecules beyond diffraction limits in cells. | ~10–50 nm. | ms–s. | Localization of protein assemblies in cells, stoichiometry. | *In vivo* context; spatial mapping. | Limited to larger assemblies; does not resolve single-protein folding pathways. |
| **Cellular and organismal models.**[230] | Manipulate and monitor proteins within living cells or organisms. | Cellular to tissue scale. | min–years. | Aggregation, toxicity, proteostasis responses. | Physiological relevance; disease modeling. | Limited molecular resolution; indirect inference of folding pathways. |

[228] K. Schwing and M. Gerhards, “Investigations on Isolated Peptides by Combined IR/UV Spectroscopy in a Molecular Beam – Structure, Aggregation, Solvation and Molecular Recognition,” *International Reviews in Physical Chemistry* 35 (2016): 569–677, https://doi.org/10.1080/0144235X.2016.1229331.
[229] C. Bustamante and S. Yan, “The Development of Single Molecule Force Spectroscopy: From Polymer Biophysics to Molecular Machines,” *Quarterly Reviews of Biophysics* 55 (2022): 9, https://doi.org/10.1017/s0033583522000087.
[230] Aleksandra E. Badaczewska-Dawid et al., “A3D Model Organism Database (A3D-MODB): A Database for Proteome Aggregation Predictions in Model Organisms,” *Nucleic Acids Research* 52 (January 2024): D360–67, https://doi.org/10.1093/nar/gkad942.

## 4. Computational Approaches Must Move Beyond Static Structure Prediction

Computational modeling has long been a central pillar of protein folding research. From early lattice models[24] to all-atom molecular dynamics simulations,[231] theory and simulation have provided mechanistic hypotheses and quantitative predictions. Recent advances in machine learning have dramatically expanded the scope of structure prediction and molecular modeling (Fig. 3). However, predicting folding dynamics presents challenges that differ fundamentally from predicting static structures. Current computational approaches span multiple levels of resolution and methodological frameworks (Table 2), but none of the existing approaches provide quantitative information of the relationship between protein sequence and folding mechanism at all-atom resolution and at scale.

### 4.1 Atomistic simulations provide mechanistic insight but remain limited in timescale

All-atom molecular dynamics (MD) simulations integrate Newton's equations of motion using empirical force fields and explicit solvent models.[232] In principle, this provides a complete mechanistic description of folding at atomic resolution. In practice, accessible timescales remain a central bottleneck, and simulations use simplified water models that sacrifice some accuracy for speed. Conventional all-atom MD simulations typically access nanoseconds to microseconds, and in exceptional cases milliseconds for small, fast-folding proteins. Millisecond-scale folding events have been captured for larger systems using specialized hardware[28] and large computational campaigns,[233] but such simulations remain computationally expensive and not broadly accessible.

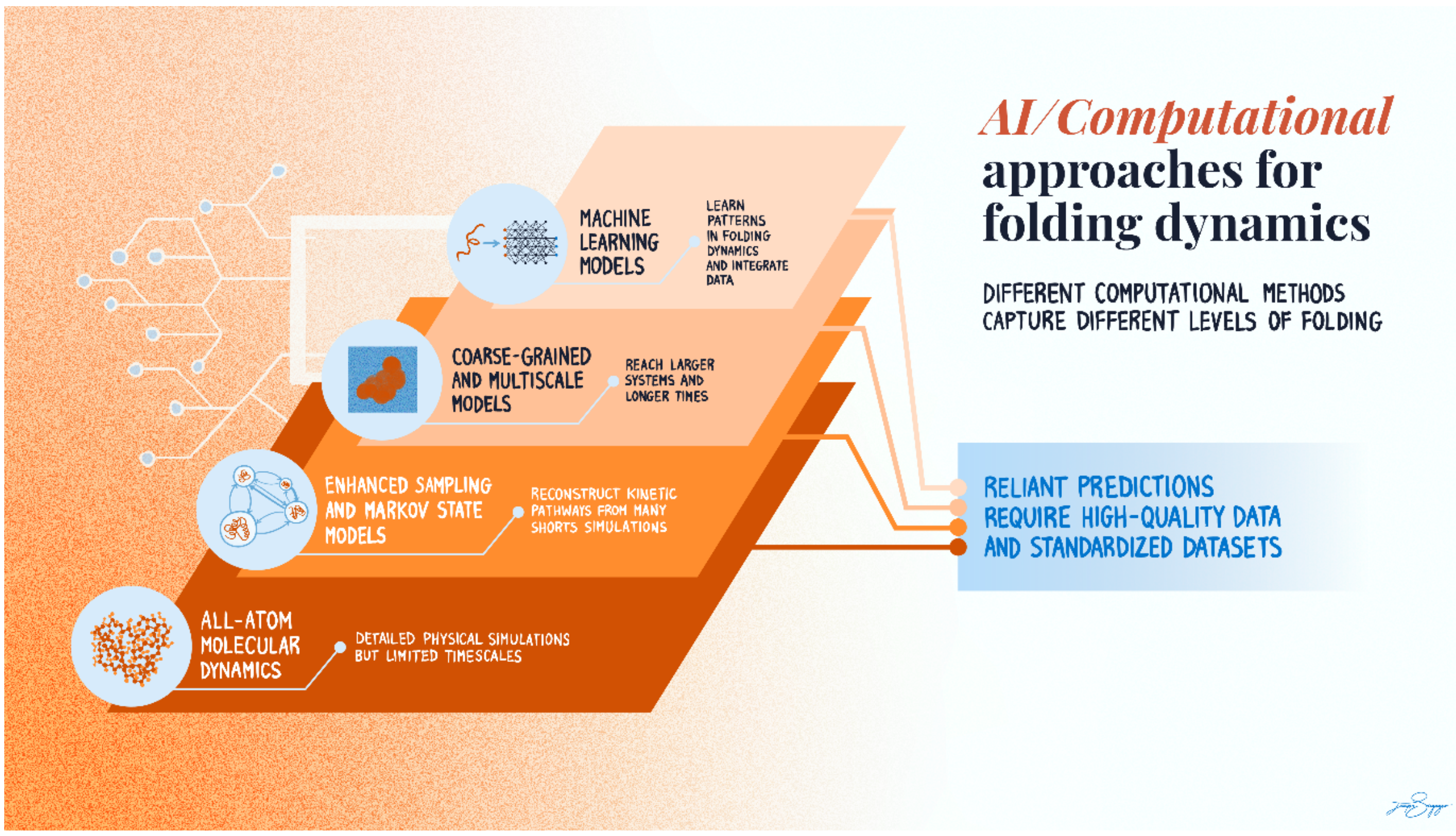

*Figure 3. **Computational advances in protein folding studies**. Current approaches address multiple levels of resolution and methodological frameworks, however, none of the existing methods provides quantitative and dynamic information of the relationship between protein sequence and folding mechanism at all-atom resolution and at scale. Graphics by J. Saquegno.*

---

[231] D. E. Shaw et al., "Atomic-Level Characterization of the Structural Dynamics of Proteins," *Science* 330 (2010): 341–46, https://doi.org/10.1126/science.1187409.
[232] M. Karplus and J. A. McCammon, "Molecular Dynamics Simulations of Biomolecules," *Nature Structural Biology* 9 (2002): 646–52, https://doi.org/10.1038/nsb0902-646.

Large distributed computing efforts have a lot of potential, especially when analyzed with Markov state models (MSMs),[234] but their capabilities are uneven across different regimes of protein behavior. First, direct observation of folding trajectories remains limited to small, fast-folding proteins, even when using specialized hardware or distributed platforms. Second, simulations are increasingly able to characterize conformational ensembles around native states (such as transitions between active and inactive conformations) for selected systems. Third, accessing misfolding and aggregation processes remains particularly challenging, as these involve larger systems and longer timescales that are often beyond current capabilities.

In fact, such large distributed efforts illustrate both the promise and the limitations of current approaches. For example, during the COVID-19 pandemic, Folding@Home generated more than 100 milliseconds of aggregate MD trajectories for SARS-CoV-2 proteins.[235] However, even at this scale, reuse for training general models proved difficult due to heterogeneous conditions, uneven sampling, and limited coverage of relevant sequence and conformational space. The challenge is therefore not only the total amount of data, but its distribution, consistency, and suitability for capturing the relevant dynamical processes.

Enhanced sampling methods such as replica exchange,[236] metadynamics,[237] and umbrella sampling,[238] extend exploration relative to standard MD simulations. These methods can reconstruct free-energy surfaces and transition networks, but formulating an effective enhanced sampling strategy is nontrivial and highly contingent on the specific characteristics of the process under investigation. Poorly chosen coordinates can bias dynamics or miss relevant pathways.[239] MSMs approximate long-timescale kinetics by stitching together multiple trajectories shorter than the timescale of the phenomenon to be observed; however, their reliability depends on sufficient state coverage and statistical convergence.[240]

---

[233] L. Casalino et al., "Beyond Shielding: The Roles of Glycans in the SARS-CoV-2 Spike Protein," *ACS Central Science* 6 (2020): 1722–34, https://doi.org/10.1021/acscentsci.0c01056.
[234] J. D. Chodera and F. Noé, "Markov State Models of Biomolecular Conformational Dynamics," *Current Opinion in Structural Biology* 25 (2014): 135–44, https://doi.org/10.1016/j.sbi.2014.04.002.
[235] M. I. Zimmerman et al., "SARS-CoV-2 Simulations Go Exascale to Predict Dramatic Spike Opening and Cryptic Pockets across the Proteome," *Nature Chemistry* 13 (2021): 651–59, https://doi.org/10.1038/s41557-021-00707-0.
[236] Y. Sugita and Y. Okamoto, "Replica-Exchange Molecular Dynamics Method for Protein Folding," *Chemical Physics Letters* 314 (1999): 141–51, https://doi.org/10.1016/S0009-2614(99)01123-9.
[237] A. Laio and M. Parrinello, "Escaping Free-Energy Minima," *Proceedings of the National Academy of Sciences* 99 (2002): 12562–66, https://doi.org/10.1073/pnas.202427399.
[238] G. M. Torrie and J. P. Valleau, "Nonphysical Sampling Distributions in Monte Carlo Free-Energy Estimation: Umbrella Sampling," *Journal of Computational Physics* 23 (1977): 187–99, https://doi.org/10.1016/0021-9991(77)90121-8.
[239] F. Noé and C. Clementi, "Collective Variables for the Study of Long-Time Kinetics from Molecular Trajectories: Theory and Methods," *Current Opinion in Structural Biology* 43 (2017): 141–47, https://doi.org/10.1016/j.sbi.2017.02.006.
[240] J. D. Chodera et al., "Automatic Discovery of Metastable States for the Construction of Markov Models of Macromolecular Conformational Dynamics," *The Journal of Chemical Physics* 126 (2007): 155101, https://doi.org/10.1063/1.2714538.

Transition path sampling (TPS) is a powerful approach designed to overcome the timescale limitations of conventional molecular simulations by focusing directly on the rare transition events that connect metastable states.[241] Instead of simulating long waiting periods between events, TPS selectively samples only the reactive trajectories, namely, paths that successfully connect two predefined stable states. This enables efficient exploration of rare processes in both all-atom and coarse-grained models. A key advantage of TPS is that it does not require prior knowledge of reaction coordinates, mechanisms, or transition states.[242] As a result, TPS has become an important tool in computational studies of protein folding, conformational transitions, and chemical reactions. However, despite its strengths, TPS faces practical challenges related to sampling efficiency. Because new trajectories are typically generated by making small modifications to existing ones (e.g., shooting or shifting moves), successive paths can become highly correlated. This reduces the effective number of independent samples in the transition path ensemble, which is crucial to compute statistical properties of the transition process, such as reaction rates, free energy profiles, and mechanistic information about the pathways connecting metastable states.

Together, these approaches provide complementary frameworks for characterizing molecular events: classical MD offers unbiased dynamical detail but is limited by timescales. Enhanced sampling techniques accelerate exploration through carefully chosen biases or collective variables. Path sampling techniques directly target the transition pathways themselves, enabling mechanistic insight without prior assumptions, but have their own sampling challenges.

### 4.2 Coarse-grained and multiscale models extend reach but require careful validation

Coarse-grained simulations reduce the number of degrees of freedom by grouping atoms into effective interaction sites.[243] This reduction allows larger time steps and results in smoother free-energy landscapes. Coarse-grained simulations can access microseconds to seconds and beyond, enabling exploration of domain assembly, oligomerization, and large conformational rearrangements.[244]

Structure-based models,[245] knowledge-based potentials,[246] and data-driven[32] or systematically derived coarse-grained force fields have been used to study folding funnels, aggregation pathways,[247] and large macromolecular complexes.[248] However, coarse-graining introduces effective interactions that are not uniquely defined. Kinetic fidelity is particularly sensitive to friction coefficients, implicit solvent representations, and mapping strategies between atomistic and reduced representations.

---

[241] L. T. Chong et al., “Path-Sampling Strategies for Simulating Rare Events in Biomolecular Systems,” *Current Opinion in Structural Biology* 43 (2017): 88–94, https://doi.org/10.1016/j.sbi.2016.11.019.
[242] P. G. Bolhuis et al., “Transition Path Sampling: Throwing Ropes over Rough Mountain Passes, in the Dark,” *Annual Review of Physical Chemistry* 53 (2002): 291–318, https://doi.org/10.1146/annurev.physchem.53.082301.113146.
[243] W. G. Noid, “Perspective: Coarse-Grained Models for Biomolecular Systems,” *The Journal of Chemical Physics* 139 (2013): 090901, https://doi.org/10.1063/1.4818908.
[244] S. J. Marrink et al., “Coarse Grained Model for Semiquantitative Lipid Simulations,” *The Journal of Physical Chemistry B* 108 (2004): 750–60, https://doi.org/10.1021/jp036508g.
[245] N. Go, “Theoretical Studies of Protein Folding,” *Annual Review of Biophysics and Bioengineering* 12 (1983): 183–210, https://doi.org/10.1146/annurev.bb.12.060183.001151.
[246] M. J. Sippl, “Knowledge-Based Potentials for Proteins,” *Current Opinion in Structural Biology* 5 (1995): 229–35, https://doi.org/10.1016/0959-440X(95)80081-6.

Multiscale frameworks aim to couple atomistic and coarse-grained levels dynamically.[249] In principle, these approaches allow transfer of information across resolutions. In practice, ensuring thermodynamic consistency and seamless coupling remain difficult. Transition rates and barrier heights can change under coarse-graining, and reweighting schemes are not always straightforward. In fact, multiscale modeling has historically struggled to link models together in a quantitatively consistent way.[250] Machine learning is increasingly viewed as a potential integrator across scales, but this integration requires principled constraints and benchmarking.[251]

Thus, coarse-grained and multiscale models extend accessible regimes but require careful validation against both atomistic simulations and experimental observables, especially when kinetic predictions are the target.

## 4.3 Machine learning can integrate scales and accelerate sampling

Machine learning is reshaping protein folding research by enhancing molecular force fields, accelerating exploration of conformational space, constructing low-dimensional representations of complex energy landscapes, and systematically integrating heterogeneous experimental and computational data.[42] Recent advances in multi-modal multiscale imaging approaches enable to connect molecular identity (specific proteins or functional sites of interest) with ultrastructural context, both at high resolution, in 3D. Such workflows enable the establishment of ground truth data in a given modality, further enabling machine-learning approaches for automation and hence high precision quantification.[252] Machine-learned interatomic potentials can learn efficient surrogates of quantum-mechanical energies and forces,[253] opening up the study of proteins with quantum-mechanical accuracy.[254] Neural samplers and generative models,[255,256] including diffusion models[77,257] and normalizing flows,[78,89] can approximate conformational distributions and sample transition pathways.[258] Simulation-based inference frameworks integrate experimental measurements with forward simulations to infer hidden states and kinetic parameters.[259] Finally, protein language and generative models learn statistical patterns in sequence space from evolutionary data.[260,261]

[247] H. D. Nguyen and C. K. Hall, "Molecular Dynamics Simulations of Spontaneous Fibril Formation by Random-Coil Peptides," *Proceedings of the National Academy of Sciences* 101 (2004): 16180–85, https://doi.org/10.1073/pnas.0407273101.
[248] A. J. Pak and G. A. Voth, "Advances in Coarse-Grained Modeling of Macromolecular Complexes," *Current Opinion in Structural Biology* 52 (2018): 119–26, https://doi.org/10.1016/j.sbi.2018.11.005.
[249] S. Izvekov and G. A. Voth, "A Multiscale Coarse-Graining Method for Biomolecular Systems," *The Journal of Physical Chemistry B* 109 (2005): 2469–73, https://doi.org/10.1021/jp044629q.
[250] M. Praprotnik et al., "Multiscale Simulation of Soft Matter: From Scale Bridging to Adaptive Resolution," *Annual Review of Physical Chemistry* 59 (2008): 545–71, https://doi.org/10.1146/annurev.physchem.59.032607.093707.
[251] G. C. Peng et al., "Multiscale Modeling Meets Machine Learning: What Can We Learn?," *Archives of Computational Methods in Engineering* 28 (2021): 1017–37, https://doi.org/10.1007/s11831-020-09405-5.
[252] C. Bosch et al., "Functional and Multiscale 3D Structural Investigation of Brain Tissue through Correlative in Vivo Physiology, Synchrotron Microtomography and Volume Electron Microscopy," *Nature Communications* 13 (2022): 2923, https://doi.org/10.1038/s41467-022-30199-6.
[253] O. T. Unke et al., "Machine Learning Force Fields," *Chemical Reviews* 121 (2021): 10142–86, https://doi.org/10.1021/acs.chemrev.0c01111.

Thus, machine learning can work as a “glue” across scales—from quantum-level calculations to coarse-grained models and experimental observables—particularly in multiscale contexts. Neural-network-based methods can, in principle, link representations across resolutions and accelerate exploration of configurational space.[262]

However, two constraints must be emphasized. First, computational scale matters. Training large foundation-style models requires substantial computing resources, often beyond what individual laboratories can sustain. Second, data curation is critical. Simply aggregating simulation data does not guarantee model performance. For example, as we have seen above, harvesting over 100 ms of publicly available MD trajectory data proved largely unusable due to inconsistent conditions and poor coverage. Effective machine-learning training requires carefully designed datasets with controlled conditions, standardized formats, and representative coverage of sequence and conformational space.

The recent success of deep learning–based structure prediction illustrates the power of curated, standardized structural databases. An equivalent, large-scale infrastructure for folding trajectories and conformational dynamics is still lacking. Without comprehensive and harmonized kinetic datasets, machine learning approaches to protein dynamics will remain fundamentally constrained by data scarcity, heterogeneity, and limited coverage of relevant timescales.

---

[254] A. Kabylda et al., “Molecular Simulations with a Pretrained Neural Network and Universal Pairwise Force Fields,” *Journal of the American Chemical Society* 147 (2025): 33723–34, https://doi.org/10.1021/jacs.5c09558.
[255] S. Olsson, “Generative Molecular Dynamics,” *Current Opinion in Structural Biology* 96 (2026): 103213, https://doi.org/10.1016/j.sbi.2025.103213.
[256] F. Noé et al., “Boltzmann Generators: Sampling Equilibrium States of Many-Body Systems with Deep Learning,” *Science* 365 (2019): 1147, https://doi.org/10.1126/science.aaw1147.
[257] M. Schreiner et al., “Implicit Transfer Operator Learning: Multiple Time-Resolution Models for Molecular Dynamics,” *Advances in Neural Information Processing Systems* 36 (2023): 34483–503, https://doi.org/10.52202/075280-1582.
[258] H. Jung et al., “Machine-Guided Path Sampling to Discover Mechanisms of Molecular Self-Organization,” *Nature Computational Science* 3 (2023): 334–45, https://doi.org/10.1038/s43588-023-00428-z.
[259] L. Dingeldein et al., “Simulation-Based Inference of Single-Molecule Experiments,” *Current Opinion in Structural Biology* 91 (2025): 102988, https://doi.org/10.1016/j.sbi.2025.102988.
[260] Z. Lin et al., “Evolutionary-Scale Prediction of Atomic-Level Protein Structure with a Language Model,” *Science* 379 (2023): 1123–30, https://doi.org/10.1126/science.ade2574.
[261] Tomas Geffner et al., “La-Proteina: Atomistic Protein Generation via Partially Latent Flow Matching,” version 2, preprint, arXiv, 2025, https://doi.org/10.48550/ARXIV.2507.09466.
[262] S. Hummerich et al., “Split-flows: Measure transport and information loss across molecular resolutions,” 2025, https://doi.org/10.48550/arXiv.2511.01464.

## 4.4 Predicting folding dynamics poses fundamentally different challenges than predicting structure

Predicting static structure from sequence is primarily a mapping from one high-dimensional space (sequence) to another (3D coordinates). Models like AlphaFold and RoseTTAFold demonstrate that, given sufficient evolutionary information and curated structural data, this mapping can be learned with high accuracy for many folded proteins.

Predicting folding dynamics is qualitatively different. The target is not a single configuration, but an ensemble of paths. A complete prediction must specify intermediate states, transition probabilities, barrier heights, and rates under defined environmental conditions. These quantities are sensitive to subtle energetic differences. Small errors in free-energy estimates can produce orders-of-magnitude differences in predicted kinetics.

A predictive model of protein folding dynamics must therefore be probabilistic, capable of representing ensembles of stationary and dynamic properties, rather than deterministic endpoints. Deep generative models (for example, diffusion models[263] and normalizing flows[264]) can learn such distributions explicitly, but it remains unclear how to train these models at scale and in a manner that transfers across chemical space and thermodynamic conditions. An important step will likely involve establishing curated datasets that must be designed with kinetic diversity, controlled perturbations, and explicit metadata describing conditions.

In summary, extending computational success from static structure to folding paths requires: (i) reliable energetics across resolutions; (ii) probabilistic models capable of representing kinetic heterogeneity, and (iii) systematically curated trajectory datasets analogous in rigor to structural databases. Table 2 summarizes the current computational landscape and clarifies that, while many methodological components exist, their integration into a predictive sequence-to-trajectory framework remains an open challenge.

---

[263] J. Ho et al., "Denoising Diffusion Probabilistic Models," *Advances in Neural Information Processing Systems* 33 (2020): 6840–51.
[264] D. Rezende and S. Mohamed, "Variational Inference with Normalizing Flows," *International Conference on Machine Learning*, June 2015, 1530–38, https://dl.acm.org/doi/10.5555/3045118.3045281.

**Table 2 | Computational and AI approaches for studying protein folding dynamics.**

| Approach | Principle | Level of Representation | Accessible Scales | What It Predicts or Calculates | Strengths | Limitations for Folding Mechanism |
|---|---|---|---|---|---|---|
| **All-atom molecular dynamics (MD).**[265] | Integrate Newton's equations of motion for all atoms using empirical force fields. | Explicit atoms and solvent. | ns–μs (rarely ms with specialized hardware). | Atomistic trajectories; folding events for small fast-folding proteins. | Physically grounded; high structural detail. | Limited timescales; computationally expensive; force-field accuracy constraints. |
| **Enhanced sampling methods (e.g., metadynamics[266] and replica exchange[267]).** | Accelerate exploration of conformational space using biasing schemes or ensemble replication. | All-atom or coarse-grained. | Extended relative to standard MD. | Free energy landscapes; rare events; transition networks. | Improved exploration of conformational space. | Sensitive to method and parameter choice, including collective variable; may bias dynamics; interpretation can be model-dependent. |
| **Path sampling.**[241,242] | Bypasses long timescales by sampling only. transition pathways. | All-atom or coarse-grained. | Rare events regardless of waiting time. | Computational studies of rare events. | Does not require prior knowledge of mechanisms, reaction coordinates, and transition states. | Struggles to efficiently and ergodically sample the full ensemble of rare transition paths, especially in high dimensions. |
| **Coarse-grained simulations.**[243] | Reduce degrees of freedom by grouping atoms into effective interaction sites. | Reduced degrees of freedom. | μs–s and beyond. | Large-scale conformational changes; assembly processes. | Increased timescales; smoother energy landscapes. | Loss of atomic detail; mapping between scales nontrivial. |
| **Multiscale modeling frameworks.**[268] | Couple models at different resolutions to exchange structural or energetic information. | Coupled atomistic and coarse-grained. | Variable. | Information transfer across resolutions. | Potential to integrate scales. | Difficult to ensure thermodynamic consistency; limited seamless coupling. |
| **Machine-learned interatomic potentials.**[253] | Learn interaction potentials from high-level simulations or experiments while enforcing physical constraints. | Atomistic with learned interactions. | Similar to MD. | Energy surfaces; forces; accelerated simulations. | Improved efficiency; learn corrections to force fields. | Requires high-quality training data; transferability challenges. |
| **Generative models for structure (e.g., AlphaFold[30] and RoseTTAFold[31]).** | Learn statistical mappings from sequence to structure using large structural datasets. | Sequence to structure mapping. | Static endpoint(s). | Native 3D structure from sequence. | High accuracy for folded states; scalable. | Does not predict folding pathways or kinetics; limited for disorder and metastability. |
| **Generative models for dynamics.**[255] | Learn probability distributions over conformations or transitions using data-driven approaches. | Variable resolution. | Model-dependent. | Distributions of conformations; transition pathways. | Flexible probabilistic modeling; can integrate data. | Limited training data for folding trajectories; physical interpretability evolving. |
| **Simulation-based inference and hybrid experimental–computational models.**[259] | Infer mechanistic parameters by matching forward simulations to experimental observables. | Model-dependent. | Model-dependent. | Parameter estimation; inference of hidden states from data. | Integrate experimental measurements with simulations. | Requires reliable forward models; computationally demanding. |
| **Protein language and structure models.**[260] | Learn statistical patterns in sequence space from evolutionary data. | Sequence-level. | Static or statistical. | Evolutionary constraints; stability signals. | Captures sequence patterns; scalable across proteomes. | Indirect link to folding kinetics; limited mechanistic interpretability. |

[265] R. O. Dror et al., "Biomolecular simulation: a computational microscope for molecular biology," *Annual Review of Biophysics* 41 (2012): 429–52, https://doi.org/10.1146/annurev-biophys-042910-155245.
[266] G. Bussi and A. Laio, "Using Metadynamics to Explore Complex Free-Energy Landscapes," *Nature Reviews Physics* 2 (2020): 200–212, https://doi.org/10.1038/s42254-020-0153-0.
[267] D. J. Earl and M. W. Deem, "Parallel Tempering: Theory, Applications, and New Perspectives," *Physical Chemistry Chemical Physics* 7 (2005): 3910–16, https://doi.org/10.1039/B509983H.
[268] G. S. Ayton et al., "Multiscale Modeling of Biomolecular Systems: In Serial and in Parallel," *Current Opinion in Structural Biology* 17 (2007): 192–98, https://doi.org/10.1016/j.sbi.2007.03.004.

# 5. Integrating Experiment and Computation Toward a Predictive Folding Framework

The preceding sections outline a structural asymmetry in the field. Static structure prediction succeeded because it combined large, curated experimental datasets with scalable machine learning architectures. Folding dynamics research possesses increasingly powerful experimental and computational tools but still lacks an equivalent integrative infrastructure. The central opportunity is to couple time-resolved experiments with multiscale modeling in a systematic and iterative manner. A predictive framework for protein folding—and more broadly for protein conformational dynamics and macromolecular self-assembly—will require coordinated development of standardized trajectory-resolved datasets, rigorous probabilistic modeling frameworks, and experimental platforms capable of generating reproducible kinetic measurements under systematically controlled perturbations.

## 5.1 The success of structure prediction illustrates the power of shared and standardized data

The success of sequence-to-structure prediction depended critically on the existence of the Protein Data Bank,[269] which provided curated, standardized, and interoperable structural data accumulated over decades. Structural entries follow common deposition formats, validation procedures, and metadata standards, allowing cross-laboratory comparability and machine-readable consistency. This infrastructure enabled machine learning systems to learn generalizable mappings from sequence to structure. The decisive factor was not simply the number of structures, but the consistency of experimental protocols and annotation standards. Thanks to this, structural models deposited from crystallography, NMR, and cryo-EM could be treated as components of a coherent dataset.

## 5.2 Folding research lacks an equivalent ecosystem of curated trajectory data

In contrast to structural databases, databases of folding trajectories and conformational dynamics are not systematically curated or standardized. Experimental kinetic measurements are often optimized for specific proteins under certain conditions. Simulation datasets vary in force fields, solvent models, temperature control, sampling strategies, and trajectory length. Metadata describing perturbations and environmental conditions are frequently incomplete or inconsistent.

The limitations of unsystematic aggregation are evident from attempts to reuse large-scale MD datasets.[270] As we have seen above, more than 100 milliseconds of aggregate MD trajectories generated during distributed computing efforts were made publicly available, yet only a small fraction proved suitable for training or benchmarking general models.[235] The principal limitations were heterogeneous simulation protocols, uneven sequence diversity, and insufficient annotation of parameters. Data volume alone did not compensate for lack of standardization. Recent efforts such as the Molecular Dynamics Data Bank (MDDB), aim to overcome these limitations by systematically archiving, standardizing, and enabling the reuse of large-scale simulation trajectories and associated metadata.[271]

---

[269] H. Berman et al., “Announcing the Worldwide Protein Data Bank,” *Nature Structural & Molecular Biology* 10 (2003): 980, https://doi.org/10.1038/nsb1203-980.
[270] J. K. Tiemann et al., “MDverse, Shedding Light on the Dark Matter of Molecular Dynamics Simulations,” *eLife* 12 (2024): 90061, https://doi.org/10.7554/eLife.90061.3.

Folding experiments face analogous challenges. Differences in buffer composition, denaturant concentration, labeling strategies, pulling geometries, and analysis pipelines complicate direct comparison of kinetic parameters across studies. Without standardized benchmarking proteins and protocols, extracting generalizable principles from heterogeneous datasets is challenging.

A curated ecosystem of folding and conformational trajectory data would require well-defined reference proteins, standardized perturbation protocols, interoperable data formats, and comprehensive metadata describing experimental conditions, interactions, and analysis pipelines. Without such infrastructure, folding kinetics remain dispersed across isolated studies, limiting comparability, reproducibility, and cumulative model development.

## 5.3 Iterative feedback between experiments and models can create a virtuous cycle

Advances in single-molecule force spectroscopy and fluorescence microscopy methods allow direct, time-resolved measurement of stochastic folding, conformational transitions, and interaction-driven assembly trajectories. [20,187,272,273] These experiments provide quantitative observables such as transition rates, barrier heights, intermediate lifetimes, and force-dependent kinetics under controlled perturbations. When acquired under standardized conditions, such measurements can constrain and validate mechanistic models.

Computational frameworks can then incorporate these constraints. Enhanced sampling simulations and Markov state models estimate kinetic networks from atomistic trajectories. Coarse-grained and multiscale models extend predictions to longer timescales. Simulation-based inference approaches adjust model parameters to reproduce experimental observables, effectively integrating measurement and theory within a probabilistic framework. Generative models can propose conformational distributions, transition pathways, and interaction-driven assembly processes consistent with empirical kinetics.

Machine learning offers a mechanism for integrating across scales by learning effective mappings between atomistic energetics, coarse-grained representations, and experimentally accessible observables. In this setting, experimental trajectories will not merely confirm simulations, but they will inform model parameterization, identify missing intermediates, and highlight discrepancies in energy barrier heights or pathway topology.

---

[271] European Commission, “Molecular Dynamics Data Bank (MDDB),” 2023, https://doi.org/10.3030/101094651.
[272] R. Roy et al., “A Practical Guide to Single-Molecule FRET,” *Nature Methods* 5 (2008): 507–16, https://doi.org/10.1038/nmeth.1208.
[273] O. K. Dudko et al., “Theory, Analysis, and Interpretation of Single-Molecule Force Spectroscopy Experiments,” *Proceedings of the National Academy of Sciences* 105 (2008): 15755–60, https://doi.org/10.1073/pnas.0806085105.

Model predictions, in turn, can guide experimental design. Predicted mutation effects on barrier heights or kinetic partitioning can be tested through targeted single-molecule measurements. Perturbations predicted to alter intermediate populations can be experimentally validated. Discrepancies can motivate refinement of force fields, coarse-graining strategies, or probabilistic architectures. Through repeated iteration, uncertainty in both experimental interpretation and model parameters can be progressively reduced.

This bidirectional exchange will establish a virtuous cycle in which trajectory-resolved experiments constrain models, models generate quantitative predictions, and new experiments test and refine those predictions. Over time, this cycle can transform studies of protein folding, conformational dynamics, and self-assembly from descriptive case analyses into cumulative, predictive science.

For this iterative cycle to operate beyond isolated case studies, the field will require shared experimental, data, and compute resources that make results comparable across proteins, instruments, laboratories, and modelling approaches. National and international Research Infrastructures can play an important role here, not as stand-alone institutional actors, but as enabling environments for reproducible measurement, data stewardship, access to advanced technologies, and cross-platform validation. High-throughput experimental platforms must generate trajectory-resolved datasets under controlled perturbations; data services must support metadata capture, quality control, versioning, and FAIR access; and compute environments must provide the storage, interoperability, and model-training capacity required for large-scale prediction.

## 5.4 A predictive framework should connect sequence to pathways and cellular outcomes

A mature predictive framework would map amino acid sequence and environmental conditions onto dynamic energy landscapes that govern conformational transitions and the formation of macromolecular assemblies. The objective would be to generate a probabilistic description of folding behavior, conformational switching, and assembly processes, that specify folding and unfolding rate constants, the relative populations and lifetimes of intermediate states, and the sensitivity of kinetic partitioning to mutation or environmental perturbation. Such a framework would also estimate protein aggregation propensity under defined physicochemical conditions and quantify how folding outcomes depend on factors such as molecular crowding or molecular chaperones interaction.

Achieving this level of prediction requires coordinated modeling across multiple spatial and temporal scales. Atomistic simulations provide detailed interaction energetics and structural resolution at the level of individual residues and solvent-mediated effects. Coarse-grained and multiscale models extend accessible timescales and enable exploration of larger assemblies, longer kinetic processes, and interaction networks underlying functional complexes. Machine learning architectures encode statistical regularities across sequence space, accelerate sampling, and provide probabilistic representations of high-dimensional conformational distributions. Experimentally measured trajectories serve as quantitative anchors, constraining model parameters and validating predicted kinetics.

Importantly, predictions must extend beyond isolated *in vitro* conditions to connect folding dynamics, conformational regulation, and assembly processes with cellular outcomes. Translation rates influence the timing and ordering of structural formation, while proteostasis capacity and degradation pathways shape the steady-state distribution of conformational states *in vivo*. Mutations may alter not only thermodynamic stability but also the balance between productive folding and off-pathway aggregation. A predictive framework should therefore couple molecular-level kinetics with cellular-scale processes that determine whether a given sequence folds efficiently, requires chaperone assistance, or is prone to accumulation of misfolded species.

From this point of view, sequence variation is interpreted through its impact on energy landscape topology and kinetic accessibility rather than solely through its effect on the stability of the native state. The shift from predicting static structures to predicting dynamic pathways and interaction-driven assembly processes is essential for understanding disease-associated misfolding, evolutionary selection pressures, and the rational engineering of folding behavior. The necessary experimental and computational components are increasingly available; their integration defines the central challenge for the next stage of protein folding research.

# 6. A Roadmap Toward Predictive Folding Dynamics and Its Impact

The field of protein folding stands at a pivotal transition, driven by advances in experimental resolution, computational modeling, and machine learning. Static structure prediction has demonstrated the power of curated data and scalable machine learning, but folding dynamics remain only partially characterized. Experimental techniques can now capture trajectory-level folding events with high spatial and temporal precision, and multiscale computational frameworks can represent complex energy landscapes. The limiting factor is now the lack of standardized trajectory datasets, probabilistic models that explicitly represent kinetic heterogeneity, and systematic, iterative coupling between experiment and simulation. This would enable sequence-to-pathway prediction, quantitative assessment of mutation-specific kinetic phenotypes, and mechanistic integration of molecular folding dynamics with cellular proteostasis networks. The next phase of protein folding research must therefore move beyond static structural endpoints toward a predictive, multiscale, and experimentally anchored framework for folding dynamics involving social sciences and humanities to ensure that emerging protein technologies are governed in socially responsible ways (Fig. 4).

## 6.1 A mechanistic understanding of folding would transform biology and medicine

A predictive framework for folding dynamics would enable direct estimation of kinetic behavior from sequence. Instead of assessing variants primarily through changes in thermodynamic stability, it would become possible to quantify how mutations alter barrier heights, intermediate populations, and the probability of misfolding under physiological conditions.

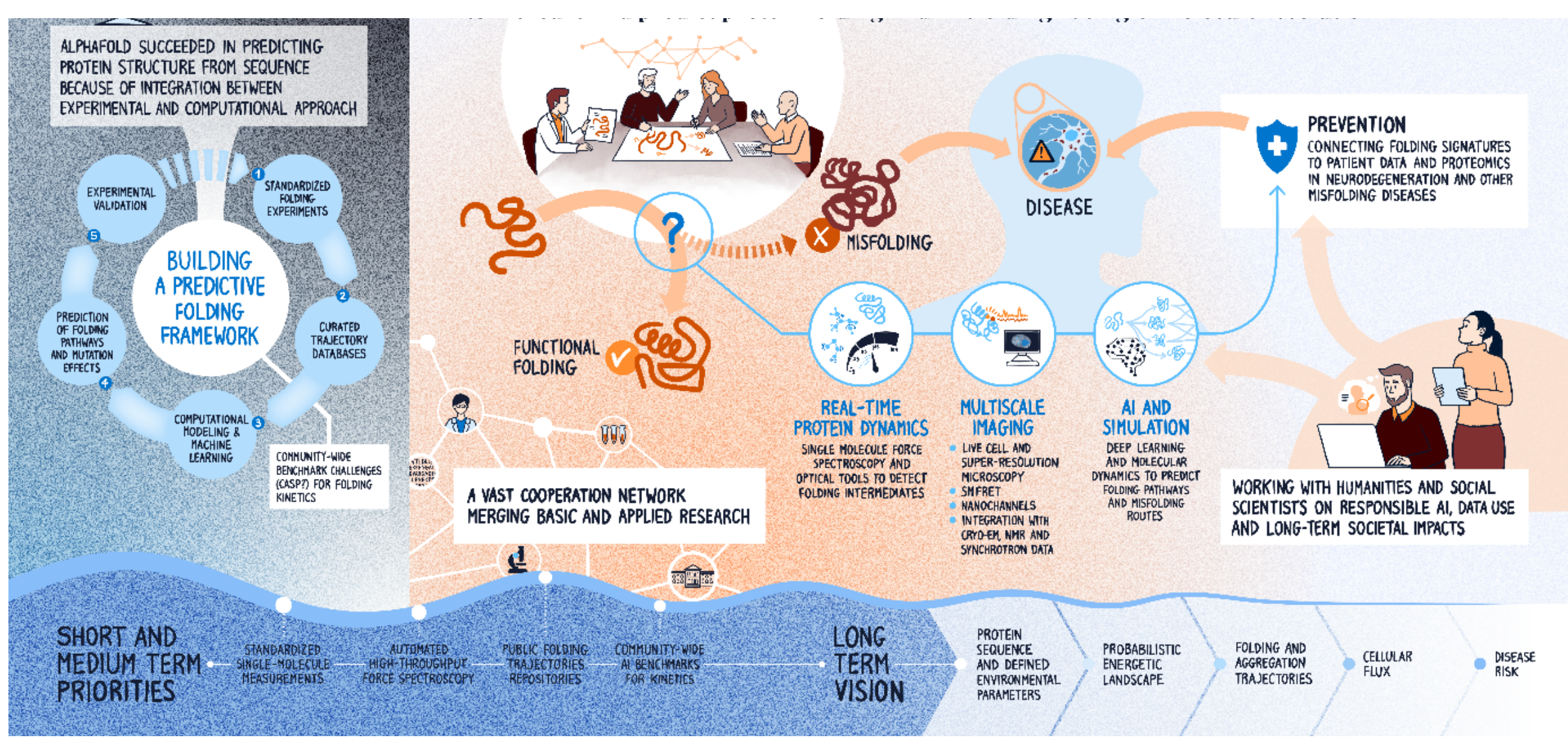

*Figure 4.* ***A roadmap towards understanding protein folding dynamics.*** *Graphics by J. Saquegno.*

In the context of disease, this would have immediate implications. Early detection of misfolding-prone protein variants could be informed by predicted kinetic signatures rather than by structural destabilization alone. Many pathogenic mutations do not abolish native structure but subtly alter folding rates or increase the lifetime of partially folded intermediates that, for example, seed aggregation. Quantifying such kinetic shifts would allow mutation-specific folding phenotypes to be defined in measurable terms.

A mechanistic description of intermediates would also open new therapeutic strategies. Recent developments in ALS illustrate the importance of mechanistic stratification. In patients carrying specific *SOD1* mutations, antibody-based therapies targeting misfolded SOD1 have demonstrated halting of clinical progression during active treatment. Notably, such effects are mutation-specific and not uniformly observed across all ALS phenotypes, underscoring the need for precise molecular subgrouping. This example suggests that defining conformationally specific and kinetically resolved disease subtypes can transform therapeutic outcomes. Rather than treating clinically heterogeneous populations as single entities, folding-aware stratification may identify small but mechanistically coherent groups in which targeted interventions are effective.

Currently, there are small molecule drugs that reshape the protein folding energy landscape, by stabilizing the native state (raising the kinetic barrier to misfolding) and/or stabilizing folding intermediates (lowering the energetic barriers to productive folding) preventing premature proteostatic degradation. A particularly successful example for therapeutics targeting protein folding states are small-molecule kinetic stabilizers of transthyretin, which have transformed the treatment of transthyretin amyloidosis patients.[274] Similarly, small-molecule correctors are used in cystic fibrosis treatment to stabilize intermediates of the CFTR[275]. In Fabry disease, such stabilizers are used to preserve native-like conformation of α-Gal A.[276] In phenylketonuria, small-molecule drugs are utilized to stabilize monomeric and tetrameric states of phenylalanine hydroxylase to improve its folding.[277] These examples serve as a proof-of-concept for the translational impact that protein folding knowledge can have on therapeutic development.

[274] P. Hammarström, "The Transthyretin Protein and Amyloidosis–an Extraordinary Chemical Biology Platform," *Israel Journal of Chemistry* 64 (2024): 202300164, https://doi.org/10.1002/ijch.202300164.
[275] G. Veit et al., "Allosteric Folding Correction of F508del and Rare CFTR Mutants by Elexacaftor-Tezacaftor-Ivacaftor (Trikafta) Combination," *JCI Insight* 5 (2020): 139983, https://doi.org/10.1172/jci.insight.139983.

Achieving predictive models of protein folding dynamics requires coordinated investment in experimental platforms, data infrastructure, and computational resources. Establishing standardized folding trajectory repositories and integrating experimental and computational workflows will likely exceed the capacity of individual laboratories. Strategic support through national and international research infrastructures could therefore play a central role in enabling this transition, similar to the role that structural biology infrastructures and the Protein Data Bank played in enabling modern structure prediction.

Integration of single-molecule measurements, AI-based models, and clinical data, would enable identification of early, pre-symptomatic changes and define patient-specific disease trajectories. This will further support earlier diagnosis, improved patient stratification, and the development of predictive tools for clinical decision-making. When aggregated or misfolded proteins drive the pathological process, these diagnostic tools can be further developed into novel therapeutics. For instance, antibodies and nanobodies can be used in diagnostics to measure protein levels and also for patient treatments, e.g., Aβ antibodies for Alzheimer's disease treatment.[278]

A comprehensive understanding of misfolding kinetics could also guide the rational design of therapeutics with novel modes of action. Rather than targeting the final aggregated state, interventions should account for the fact that, in many amyloid-related diseases, the most toxic species are transient intermediate aggregates rather than mature fibrils. Therapeutic strategies could therefore aim to modulate the formation, stability, or clearance of these intermediates, for example by stabilizing non-pathological states, accelerating productive folding pathways, or diverting aggregation toward less toxic end states. Small molecules or chaperone modulators should thus be evaluated not only for binding affinity, but also for their effects on kinetic partitioning between folding, intermediate states, and aggregation routes.

More broadly, a predictive framework would enable rational engineering of folding kinetics in biotechnology. Construct design of recombinant proteins could be directed not only for structural stability, but also for defined folding pathways compatible with cellular expression systems. Such effort would increase the yield and success rate of these efforts, potentially also enabling expression of proteins which are currently inaccessible to a recombinant setting. Synthetic biology applications could incorporate folding-rate constraints into sequence design pipelines.

---

[276] Xi Li et al., "Fabry Disease: Mechanism and Therapeutics Strategies," *Frontiers in Pharmacology* 13 (October 2022): 1025740, https://doi.org/10.3389/fphar.2022.1025740.
[277] C. Harding, "New Era in Treatment for Phenylketonuria: Pharmacologic Therapy with Sapropterin Dihydrochloride," *Biologics* 4 (2010): 231–36, https://doi.org/10.2147/BTT.S3015.
[278] E. Stögmann and R. Schmidt, "Amyloid-beta antibody treatment in Alzheimer's disease," *Wiener klinische Wochenschrift* 137 (2025): 182–88, https://doi.org/10.1007/s00508-024-02466-7.

## 6.2 Short- and medium-term goals should focus on standardization and benchmarking

The foremost short- and medium-term priority is the coordinated generation of standardized, high-quality kinetic datasets that can serve as robust benchmarks for predictive models of folding dynamics. While this is a matter of acquiring more measurements, it also requires agreed experimental protocols, well-characterized protein systems, shared metadata standards, reproducible analysis pipelines, and objective evaluation frameworks that allow experiments and models to be compared across laboratories. National and international research Infrastructures can contribute to this transition by providing continuity, technical expertise, data stewardship, and access to advanced platforms.

First, there is a need to scale and standardize single-molecule measurements. Single-molecule force spectroscopy, including optical and magnetic tweezers, already provides direct measurements of barrier heights, transition rates, intermediate states, and force-dependent kinetics. However, systematic scaling is required if these measurements are to become broadly usable for model training and validation. Instrumentation should achieve improved force stability and sub-nanometer extension resolution to resolve short-lived intermediates. Automation of tether formation, alignment, calibration, and state detection will be essential to reduce operator bias and improve reproducibility. Multiplexed configurations capable of measuring multiple molecules in parallel should become more widely available, enabling statistically robust datasets across defined perturbation regimes.

In parallel, *in vitro* smFRET-based folding assays should evolve into scalable and standardized platforms in which environmental parameters can be systematically controlled and varied. Unlike traditional measurements performed under fixed conditions, such platforms would enable the generation of trajectory-resolved datasets across well-defined perturbation regimes, making it possible to map how folding kinetics, conformational transitions, and assembly processes depend on physicochemical context. Shared technology platforms can accelerate this transition by integrating sample preparation, measurement, calibration, and analysis into reproducible workflows compatible with common data and metadata standards.

Second, protein production needs to be harmonized for applications focused on folding and conformational dynamics. Reliable and standardized protein production pipelines are a prerequisite for scaling kinetic measurements across diverse sequence classes. Many relevant systems, including engineered variants, truncated domains, membrane proteins, intrinsically disordered regions, and *de novo*-designed proteins, are difficult to express, label, purify, and validate reproducibly. Although protein production facilities are widely available, their workflows are not always optimized for quantitative folding kinetics or for direct compatibility with single-molecule and high-throughput dynamic measurements. Systematic variation in yield, solubility, stability, labelling efficiency, and construct behavior should be taken into account, because these parameters can themselves provide information about folding propensity, domain stability, and sequence-dependent behavior. Harmonized production workflows should therefore include construct design, labelling strategies, quality-control benchmarks, purification protocols, and biophysical validation under comparable conditions. Establishing such pipelines would enable systematic exploration of engineered variants, truncated domains, membrane proteins, intrinsically disordered regions, and de novo-designed sequences. This would move folding research beyond reliance on a limited set of tractable model systems toward scalable interrogation of sequence space.

Third, perturbation protocols and metadata need to be standardized. Folding and unfolding rates should be measured across defined force ranges, temperature intervals, buffer compositions, solvent conditions, crowding regimes, redox states, and other relevant environmental variables. Metadata describing construct design, expression system, labelling strategy, pulling geometry, perturbation history, instrument calibration, and analysis pipeline should be recorded in machine-readable formats. Such harmonization is essential for quantitative comparison across laboratories, instruments, protein-expression systems, and modelling approaches. Scalable *in vitro* folding platforms provide a natural framework for implementing standardized perturbation protocols in a reproducible and machine-readable manner. Complementary high-throughput approaches can broaden coverage at lower temporal resolution. For example, smFRET can quantify conformational distributions and transition kinetics across multiple constructs in parallel, while mutational scanning combined with folding-sensitive reporters can map sequence–folding relationships at scale. Although these approaches may be temporally coarser than high-resolution force spectroscopy, they expand coverage across sequence space and help identify classes of kinetic behavior, mutation effects, and environmental dependencies that can then be examined in greater mechanistic detail.

Fourth, the field needs public folding-trajectory repositories and benchmark datasets. These resources should store raw time traces, processed kinetic parameters, analysis outputs, uncertainty estimates, and complete metadata under unified reporting standards. Benchmark protein panels should span diverse topologies, sizes, folding rates, sequence classes, and biological contexts, and should be accompanied by clearly specified perturbation regimes and reporting conventions. Such panels would play a role analogous to reference datasets in structure prediction and molecular simulation, but focused on folding trajectories, kinetic observables, conformational transitions, and intermediate populations. Sustained data stewardship will be essential. Long-term governance, quality control, versioning, persistent identifiers, data-access policies, and interoperability with modelling platforms are all required if folding-trajectory datasets are to support cumulative model development.

Fifth, the field needs benchmarks for kinetic prediction. Just as CASP, the Critical Assessment of Structure Prediction,[279] , provided blind evaluation for protein structure prediction, community-wide challenges could assess the prediction of folding rates, barrier heights, intermediate populations, conformational-state distributions, and perturbation-dependent kinetic changes under predefined experimental conditions. Models should be required to predict quantitative kinetic observables, together with uncertainty estimates, before experimental results are disclosed. Such benchmarks would provide objective metrics for progress and would clarify which modelling strategies generalize beyond individual case studies.

[279] A. Kryshtafovych et al., “Critical Assessment of Methods of Protein Structure Prediction (CASP)—Round XIII,” *Proteins* 87 (2019): 1011–20, https://doi.org/10.1002/prot.25823.

## 6.3 Long-term efforts should aim to predict folding pathways directly from sequence

The long-term objective should be a sequence-to-pathway predictive framework that integrates multiscale modeling with experimental kinetic data. In this framework, a protein sequence, together with defined environmental parameters (for example, temperature, ionic strength, translation rate, molecular crowding, and chaperone availability) would specify a probabilistic energy landscape. From this landscape, the model would estimate folding and unfolding rate constants, identify dominant and minor pathways, quantify intermediate lifetimes, and evaluate the probability of accessing aggregation-prone conformations.

Crucially, such predictions must extend beyond intrinsic folding in dilute solution. Many disease-associated proteins do not fail to fold entirely; rather, they display subtle kinetic shifts that increase the lifetime of partially folded intermediates or alter the balance between productive folding and off-pathway aggregation. A predictive model should therefore estimate kinetic partitioning under physiologically relevant conditions and quantify how small changes in barrier height or intermediate stability affect long-term conformational flux.

In neurodegenerative diseases, for example, pathology often develops over decades. Misfolded or aggregated species accumulate gradually and differentially depending on the molecular defect they contain,[280] and clinical symptoms emerge only after prolonged imbalance between production, folding, and clearance. Animal and cellular models necessarily compress these timescales, modeling within weeks or months processes that in humans unfold over years. A quantitative kinetic framework would allow extrapolation across scales by linking molecular transition rates to accumulation dynamics under defined synthesis and clearance rates. Such integration requires coupling molecular folding models to proteostasis networks. The effective concentration of misfolded intermediates depends on intrinsic folding kinetics as well as on translation rates, chaperone machinery type and availability, degradation efficiency, and cell-type- and condition-specific clearance mechanisms such as microglial activity in the brain. Variability in clearance capacity, in particular, may explain differences between species and between individuals pointing towards adaptation strategies to altering environments.

By embedding kinetic measurements within interoperable clinical and systems-level datasets, folding dynamics could be linked to genomic variation, biomarker profiles, and longitudinal disease trajectories. This integration would enable incorporation of folding kinetics into patient stratification frameworks, risk prediction models, and early-intervention strategies, this transforming protein folding from a mechanistic descriptor into a clinically actionable parameter.

Clinically, this would enable mutation-specific folding phenotypes to be defined quantitatively. Beyond mutation classification, such a framework could redefine disease staging itself. Current staging approaches in disorders such as Alzheimer's disease rely heavily on binary biomarker thresholds. However, pathological protein levels do not map linearly onto clinical decline. Some individuals with high biomarker burden remain stable, whereas others progress rapidly. A folding-informed staging system would incorporate conformational specificity, intermediate population dynamics, and predicted kinetic fragility. This would allow distinction between stable molecular states and trajectories likely to evolve toward pathological aggregation, supporting earlier, but more selective, intervention. Instead of classifying variants primarily as destabilizing or benign based on static structural models, variants could be characterized by predicted changes in folding rate, intermediate lifetime, aggregation nucleation probability, and sensitivity to chaperone modulation as exemplified in the case of transthyretin amyloidosis.[281] Such kinetic fingerprints could inform early risk assessment before overt aggregation is detectable by imaging or fluid biomarkers.

[280] D. Lumkwana et al., "Autophagic Flux Control in Neurodegeneration: Progress and Precision Targeting—Where Do We Stand?," *Progress in Neurobiology* 153 (2017): 64–85, https://doi.org/10.1016/j.pneurobio.2017.03.006.

Early detection strategies could benefit from identifying kinetic signatures that precede large-scale aggregation. For example, if a variant is predicted to increase the steady-state population of a specific partly folded intermediate, targeted assays could be developed to monitor that state in patient-derived cells. Similarly, therapeutic interventions could be evaluated for their capacity to alter kinetic partitioning rather than solely their affinity for aggregated end states. Small molecules, chaperone modulators, or translation-rate regulators could be assessed quantitatively for their impact on barrier heights and intermediate stabilization.

In protein engineering and biotechnology, sequence-to-pathway prediction would allow rational control of folding to optimize expression yields, reduce aggregation during manufacturing, and tailor properties for specific applications. Synthetic constructs could be screened *in silico* for both structural compatibility and kinetic robustness under cellular expression constraints.

Achieving this vision requires sustained integration across measurement and modeling. High-resolution force spectroscopy datasets, improved in sensitivity and throughput, will define barrier heights and intermediate states for benchmark systems. Complementary large-scale measurements (such as FRET-based conformational profiling and mutational scanning with folding-sensitive reporters) will expand coverage across sequence space. Curated trajectory repositories and standardized kinetic benchmarks will provide training and validation data for probabilistic, multiscale models. These models, in turn, will be iteratively refined through targeted experimental perturbations.

Ultimately, prevention may represent the most profound application of predictive folding dynamics. If early kinetic signatures of vulnerability can be identified before irreversible aggregation or neuronal loss occurs, interventions could aim to rebalance folding landscapes rather than remove established aggregates. This shift, from clearing pathological end states to stabilizing productive folding pathways, would mark a conceptual transition from reactive treatment to proactive maintenance of proteostasis.

# Acknowledgements

This Perspective emerged from a series of workshops and discussions on the future of protein folding research organized at the University of Gothenburg, Karolinska Institute, and Chalmers University of Technology in 2026, including meetings on *Methods and Translation* and *AI and Simulation*, focused seminars and roundtable discussions with invited international experts, and the international *Protein Folding Conference* (Stockholm, March 2026). These events were supported by the Swedish Research Council (Vetenskapsrådet) through project #2025-07534. We thank Tove Sjögren, Anders Hogner, Christian Tyrchan, and Leonardo de Maria from AstraZeneca for insightful discussions. We thank Professor Harm H Kampinga from the University of Groningen and Benjamin Schuler from the University of Zurich for critical reading and suggestions on the manuscript. We also thank Agnese Callegari from the Soft Matter Lab at the University of Gothenburg and Aykut Argun from Vintergatan Photography for their valuable support.

---

[281] Y. Sekijima et al., “The Biological and Chemical Basis for Tissue-Selective Amyloid Disease,” *Cell* 121 (2005): 73–85, https://doi.org/10.1016/j.cell.2005.01.018.